\documentclass[aps,prb,superscriptaddress,floatfix,twocolumn]{revtex4-2}

\usepackage{graphicx,amssymb,bm,amsfonts,amsmath,graphics,color,epstopdf,verbatim}
\usepackage[bookmarks=false,pdfstartview=FitH,colorlinks=true,citecolor=blue,linkcolor=blue]{hyperref}
\usepackage{natbib}
\usepackage{floatrow}
\usepackage{nameref}
\newcommand{\ket}[1]{|#1\rangle}
\newcommand{\bra}[1]{\langle#1|}

\begin{document}
\title{New three-dimensional dispersion in the type-II Dirac semimetals PtTe$_2$ and PdTe$_2$ revealed through Angle Resolved Photoemission Spectroscopy}

\author{Ivan Pelayo}
\affiliation{Department of Physics and Astronomy, California State University Long Beach, Long Beach, California 90840, USA}
\altaffiliation{IP and DB contributed equally to this work.}

\author{Derek Bergner}
\affiliation{Department of Physics and Astronomy, California State University Long Beach, Long Beach, California 90840, USA}
\altaffiliation{IP and DB contributed equally to this work.}

\author{Archibald J. Williams}
\affiliation{Department of Chemistry and Biochemistry, The Ohio State University, Columbus, Ohio 43210, USA}

\author{Jiayuwen Qi}
\affiliation{Department of Materials Science and Engineering, The Ohio State University, Columbus, Ohio 43210, USA}

\author{Penghao Zhu}
\affiliation{Department of Physics, The Ohio State University, Columbus, Ohio 43210, USA}

\author{Mahfuzun Nabi}
\affiliation{Department of Physics and Astronomy, California State University Long Beach, Long Beach, California 90840, USA}
\altaffiliation{IP and DB contributed equally to this work.}

\author{Warren L. B. Huey}
\affiliation{Department of Chemistry and Biochemistry, The Ohio State University, Columbus, Ohio 43210, USA}

\author{Luca Moreschini}
\affiliation{Department of Physics, University of California Berkeley, California 94720, USA}
\affiliation{Materials Sciences Division, Lawrence Berkeley National Laboratory, Berkeley, California 94720, United States}

\author{Ziling Deng}
\affiliation{Department of Materials Science and Engineering, The Ohio State University, Columbus, Ohio 43210, USA}

\author{Jonathan Denlinger}
\affiliation{Advanced Light Source, Lawrence Berkeley National Laboratory, Berkeley, California 94720, USA}

\author{Alessandra Lanzara}
\affiliation{Department of Physics, University of California Berkeley, California 94720, USA}
\affiliation{Materials Sciences Division, Lawrence Berkeley National Laboratory, Berkeley, California 94720, United States}

\author{Yuan-Ming Lu}
\affiliation{Department of Physics, The Ohio State University, Columbus, Ohio 43210, USA}

\author{Wolfgang Windl}
\affiliation{Department of Materials Science and Engineering, The Ohio State University, Columbus, Ohio 43210, USA}

\author{Joshua Goldberger}
\affiliation{Department of Chemistry and Biochemistry, The Ohio State University, Columbus, Ohio 43210, USA}

\author{Claudia Ojeda-Aristizabal}
\affiliation{Department of Physics and Astronomy, California State University Long Beach, Long Beach, California 90840, USA}
\email[Corresponding author~]{Claudia.Ojeda-Aristizabal@csulb.edu}
\email{Claudia.Ojeda-Aristizabal@csulb.edu}

\date{\today}

\begin{abstract}
PtTe$_2$ and PdTe$_2$ are among the first transition metal dichalcogenides that were predicted to host type-II Dirac fermions,
exotic particles prohibited in free space. These materials are layered and air-stable, which makes them top candidates for  technological applications that take advantage of their anisotropic magnetotransport properties. Here, we provide a detailed characterization of the electronic structure of PtTe$_2$ and PdTe$_2$ using Angle Resolved Photoemission Spectroscopy (ARPES) and Density Functional Theory (DFT) calculations, unveiling a new three-dimensional dispersion in these materials. Through the use of circularly polarized light, we report a different behavior of such dispersion in PdTe$_2$ compared to PtTe$_2$, that we relate to a symmetry analysis of the dipole matrix element. Such analysis reveals a link between the observed circular dichroism and the different momentum-dependent terms in the dispersion of these two compounds, despite their close similarity in crystal structure. Additionally, our data shows a clear difference in the circular dichroic signal for the type-II Dirac cones characteristic of these materials, compared to their topologically protected surface states. 
Our work provides a useful reference for the ARPES characterization of other transition metal dichalcogenides with topological properties and illustrates the use of circular dichroism as a guide to identify the topological character of
two otherwise equivalent band dispersions, and to recognize different attributes in the band structure of similar materials.  
\end{abstract}



\maketitle

\section{Introduction}
PtTe$_2$ and PdTe$_2$ are known examples of topological semimetals created via a band inversion, with four-fold degenerate band touchings that are protected by time reversal $T$ and inversion $P$ symmetries. These band touchings are fourfold-degenerate Dirac points stabilized at the same crystal momenta by a C$_3$ symmetry along the rotation axis, 
and occur away from the time-reversal-invariant momentum (TRIM) \cite{Yang2014}.
In PtTe$_2$ and PdTe$_2$, these Dirac points host type-II excitations, that are only possible in a solid and have no counterpart in free space as they violate Lorentz invariance \cite{RevModPhys.93.025002}; they were first predicted for certain Weyl semimetals, Dirac analogous where either $T$ or $P$ symmetries are broken \cite{PhysRevLett.115.265304, soluyanov_type-ii_2015}, and were later predicted to appear in transition metal dichalcogenides such as PtTe$_2$ and PdTe$_2$ \cite{PhysRevB.94.121117}.

In general, materials with $T$ and $P$ symmetries and nonsymmorphic space groups (defined by the presence of either glide plane or screw axis symmetry) are characterized by Dirac nodes located at high symmetry points at the boundaries of the first Brillouin zone \cite{Gao_Topological_2019, PhysRevLett.108.140405}. In those materials, the presence of a Dirac cone depends on the nonsymmorphic character of the crystal space group, and it is known as an essential or symmetry-enforced band crossing. In contrast, in materials with symmorphic space groups such as PtTe$_2$ and PdTe$_2$, a Dirac cone results from a band inversion that is protected by the symmorphic crystal symmetries, such as C$_3$. They are less stable than the symmetry enforced Dirac cones, as they can dissappear after turning external parameters that don't compromise the crystal symmetry, \cite{yang_classification_2014,yang_topological_2015,Gao_Topological_2019} and they are known as accidental band crossings \cite{RevModPhys.93.025002}. In these materials, type-II fourfold degenerate Dirac points are formed therefore by two twofold degenerate inverted bands that have different rotation eigenvalues. 
This was also the case in the first predicted \cite{wang_dirac_2012,wang_three-dimensional_2013} and later experimentally realized type-I Dirac semimetals, Na$_3$Bi \cite{liu_discovery_2014} and Cd$_3$As$_2$ \cite{neupane_observation_2014,borisenko_experimental_2014,liu_stable_2014}. As a result of $T$ and $P$ symmetries, the type-II Dirac dispersions in PtTe$_2$ and PdTe$_2$ come in pairs at points in momentum space along the rotation axis, away from high-symmetry points such as $\Gamma$ or A. 

The experimental signature of Lorenz-violating type-II Dirac excitations in PtTe$_2$ and PdTe$_2$ 
was first obtained through ARPES experiments \cite{Yan2017, PhysRevLett.119.016401}, supported by first principles calculations. Interestingly, these and previous experiments \cite{Liu_2015} found evidence that these materials have, in addition to the type-II Dirac dispersions, topologically non-trivial surface states. Such states derive from band inversions and connect gapped bulk bands. 

Although both three-dimensional Dirac dispersions and surface state cones arise from band inversion, their properties are fundamentally different. For instance, the three-dimensional type-II Dirac cone is associated to gapless matter, and it is formed by a four-band crossing leading to the previously mentioned fourfold degenerate Dirac point preserved by the crystal's C$_3$ rotational symmetry. It is not topologically protected as the associated topological invariant, the total Chern number of the bands, is zero \cite{moessner_moore_2021}. 
Indeed, the type-II Dirac cone results from the merging of Weyl points at opposite momenta with opposite Chern numbers. 
In contrast, the Dirac surface states are typically associated to three-dimensional bulk insulators with important spin-orbit coupling and time-reversal symmetry. Such surface states are metallic, chiral (the spin of the electrons is linked to their direction of motion) and most importantly, are topologically protected.

The different nature of these linear dispersions in Dirac semimetals manifests experimentally in different ways. In ARPES, the topological protected surface states present a switching of the circular dichroic signal across the Dirac point, related to their chiral character. In contrast, the three-dimensional type-II cones present no switching, as they are formed by twofold degenerate bands. 
Here we present an analysis of the band structure of PtTe$_2$ and PdTe$_2$ through ARPES and density functional theory (DFT) calculations that show proof of an additional dispersion that presents no switching in the circular dichroism for PtTe$_2$ and a clear switching for PdTe$_2$.
We understand our observation through a symmetry analysis of the dipole matrix elements that relate to the observed ARPES intensity, showing that even though these materials belong to the same space group, they present a dispersion that is dominated by a linear term in momentum in the case of PtTe$_2$ and a cubic term in the case of PdTe$_2$. Additionally, our photoemission data using circularly polarized light present further evidence of the different character of the known type-II Dirac cones and topologically protected surface states in PtTe$_2$ and PdTe$_2$.

\section{Methods}

Single crystals of PtTe$_2$ were grown from a flux of 1:17 ratio of Pt powder (99.98\%, Alfa Aesar) and Te chunks (99.9999\%, Alfa Aesar). This elemental ratio was placed in a fused silica quartz ampoule followed by the addition by quartz wool. This tube was sealed under a $\leq$ 60 mTorr Ar atmosphere and heated to 1000 $^{\circ}$C over 8 h, held at this temperature for additional 8 h, and then cooled at 2 $^{\circ}$C/h to 500 $^{\circ}$C. At 500 $^{\circ}$C, the tube was removed from the furnace, inverted, and centrifuged at 1000 rpm for 10 min for excess Te removal. X-ray diffraction (XRD) experiments demonstrated that PtTe$_2$ crystallizes into a one-layer trigonal unit cell with a P$\overline{3}m1$ space group. The measured lattice constants were a = b = 4.025 $\text{\AA}$ and c = 5.224 \AA \cite{doi:10.1021/acsnano.1c08681}. Single crystals of PdTe$_2$ were grown from a stoichiometric melt of Pd powder (99.95\%, Strem) and Te lump (99.999\%, Thermo Scientific). The elemental precursors were added to a fused silica quartz ampoule and sealed under a $\leq$ 60 mTorr Ar atmosphere. The tube was heated to 1000 $^{\circ}$C over 8 h, held at this temperature for additional 8 h, and then cooled at 2 $^{\circ}$C/h to 500 $^{\circ}$C, the furnace was then turned off and the tube was allowed to cool to room temperature. X-ray diffraction (XRD) experiments demonstrated that PdTe$_2$ crystallizes into a one-layer trigonal unit cell with a P$\overline{3}m1$ space group. The measured lattice constants are a = b = 4.037 $\text{\AA}$ and c = 5.130 \AA \cite{doi:10.1021/acsnano.1c08681}.
High-resolution ARPES experiments were performed at
Beamline 4.0.3 (MERLIN) of the Advanced Light Source at the Lawrence Berkeley National Lab in a vacuum better than 5$\times$10$^{-11}$ Torr using 30 - 128 eV linearly and circularly polarized photons. The total-energy resolution was 20 meV with an angular resolution $\Delta \theta$ $\leq$ 0.2$^{\circ}$. 
The density functional theory calculations were performed on $\mathrm{PtTe_2}$ and $\mathrm{PdTe_2}$ primitive cells using the Vienna Ab initio Simulation Package (VASP) \cite{kresse1993ab,kresse1994ab} with Perdew-Burke-Ernzerhof (PBE) functionals \cite{blochl1994projector,perdew1996generalized}. The relaxed electronic configurations were obtained using a $4\times4\times3$ k-point grid and a plane-wave energy cutoff of 220 eV. The lattice parameters were fixed to $a=b=4.0259$\r{A} and $c=5.2209$\r{A} for $\mathrm{PtTe_2}$ \cite{kliche1985far}; $a=b=4.0420$\r{A} and $c=5.1253$\r{A} for $\mathrm{PdTe_2}$ \cite{hoffman1976phase}. Spin-orbit coupling was included in both cases.

\begin{figure}[t]
\includegraphics[width=8cm]{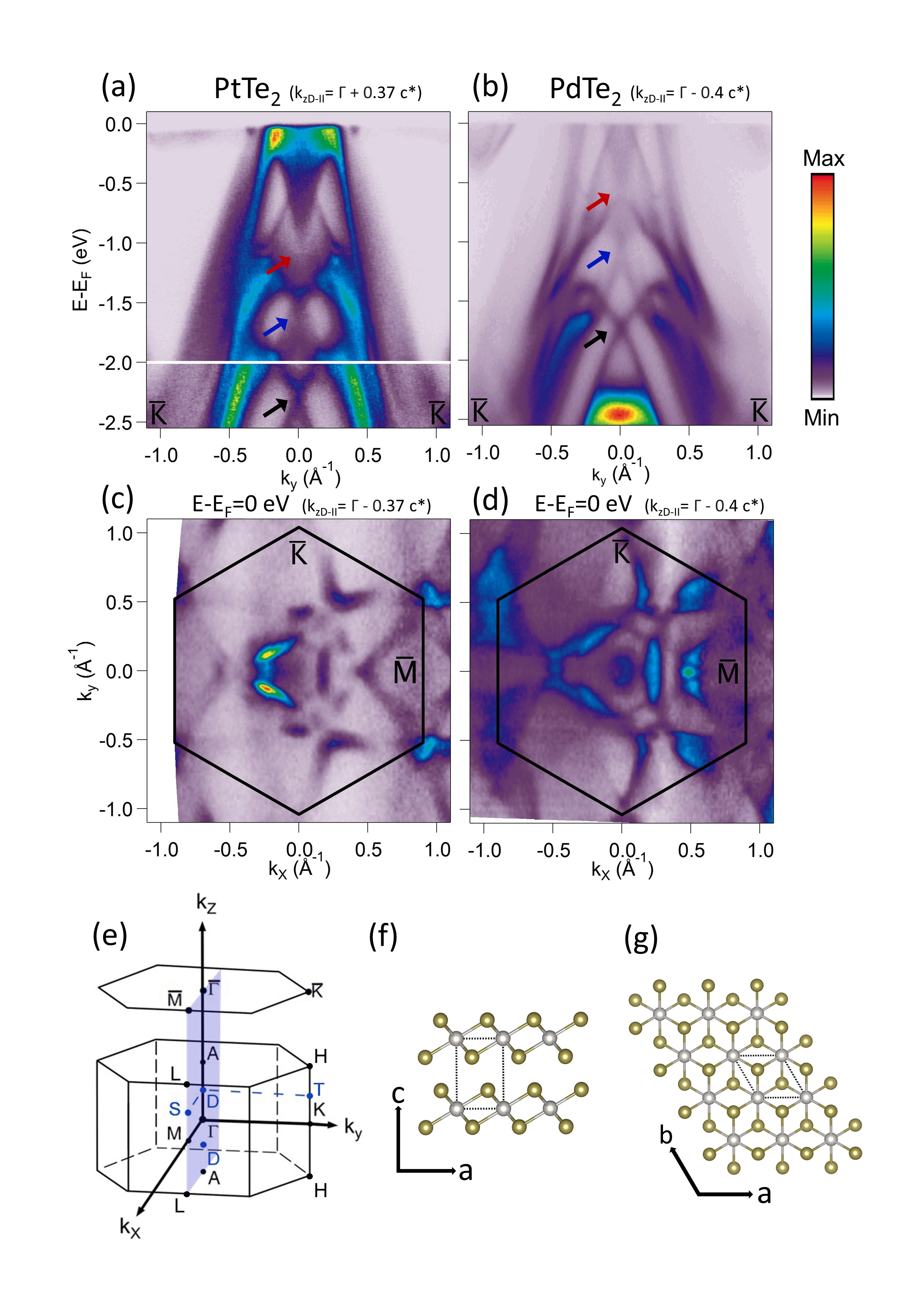}
\caption{\label{83LH}PtTe$_2$ and PdTe$_2$ high resolution ARPES band structure and Fermi surface using linearly p-polarized photons. (a) Data for PtTe$_2$ at $h\nu$ = 98 eV, the top type-II Dirac point D (k$_z$=4.37 c*) along the $\overline{\text{K}}$-$\overline{\Gamma}$-$\overline{\text{K}}$ direction showing the type-II cone (red arrow), surface state cone (black arrow) and the new 3D Dirac-like dispersion (blue arrow). (b) Data for PdTe$_2$ at $h\nu$ = 60 eV, very close to the bottom type-II Dirac cone D (k$_z$=3.55 c*) along the $\overline{\text{K}}$-$\overline{\Gamma}$-$\overline{\text{K}}$ direction. Arrows show the same Dirac dispersions as in (a). Fermi surface for PtTe$_2$ (c) and PdTe$_2$ (d) at $h\nu$ = 68 eV, close to the bottom type-II Dirac point D (k$_z$=3.63 c*) and $h\nu$ = 60 eV (k$_z$=3.55 c*, close to the bottom D k$_z$=3.60 c*) respectively.
(e) First Brillouin zone for both compounds showing the pair of type-II Dirac points and one of the three mirror planes along M-$\Gamma$-A. (f) Crystal structure for both compounds showing the transition metal atoms (white) sandwiched in-between the Te atoms (gold). (g) In-plane crystal structure for both compounds.
}
\end{figure}

\section{ARPES measurements and DFT calculations}

Fig. \ref{83LH}(e), (f) and (g) show the crystal structure and first Brillouin zone 
of the two compounds studied in this paper, PtTe$_2$ and PdTe$_2$. These compounds, similarly to PtSe$_2$ and PtBi$_2$, crystallize into space-group P$\bar{3}$m1 (No 164). They are layered, with Te atoms that sandwich the Pt (or Pd) atoms and with the Te atoms within each layer related by inversion symmetry \cite{PhysRevB.94.121117}. Space group P$\bar{3}$m1 has a primitive Bravais lattice with inversion symmetric three-fold rotation and mirror symmetry. In such a unit cell (see Fig. S1 Supp. Mat.), there are three mirror planes that coincide with the M-$\Gamma$-A planes in momentum space, highligthed in blue in Fig. \ref{83LH}(e). From these symmetries derive the main features of interest in PtTe$_2$ and PdTe$_2$, indicated by arrows in 
Fig. \ref{83LH}(a) and (b) namely, 1) a type-II Dirac cone at -1 eV and -0.6 eV for PtTe$_2$ and PdTe$_2$ respectively (red arrows),
 2) conical dispersions at -2.25 eV for PtTe$_2$ and at -1.7 eV for PdTe$_2$ that correspond to surface state Dirac cones (black arrows) and 3) a new Dirac-like dispersion at -1.4 eV for PtTe$_2$ and -1.1 eV for PdTe$_2$ (blue arrows). The data presented in Fig. \ref{83LH} were taken at photon energies close to the type-II Dirac cone D, located at k$_z$=$\Gamma\pm$0.37$c*$ for PtTe$_2$ and k$_z$=$\Gamma\pm$0.40$c*$ for PdTe$_2$. For PtTe$_2$ we used $h \nu$=68 eV, close to k$_z$=3.63c* as well as $h \nu$=98 eV, close to k$_z$=4.37c*.  
We considered an inner potential of 
V$_0$=12 eV, with c$^*$=$2\pi$/c and c=5.224 \AA, deduced from XRD experiments. In the case of PdTe$_2$, data were collected using $h \nu$=60 eV, close to k$_z$=3.60c*
assuming an inner potential of 
V$_0$=16 eV with c=5.130 \AA (see tables I and II in the Supp. Mat.). In both cases, we used p-polarized photons. Fig. \ref{83LH} (c) and (d) show constant energy maps at the Fermi level for PtTe$_2$ and PdTe$_2$ respectively, where we can observe a three-fold symmetry.

\begin{figure}
\includegraphics[width=9cm]{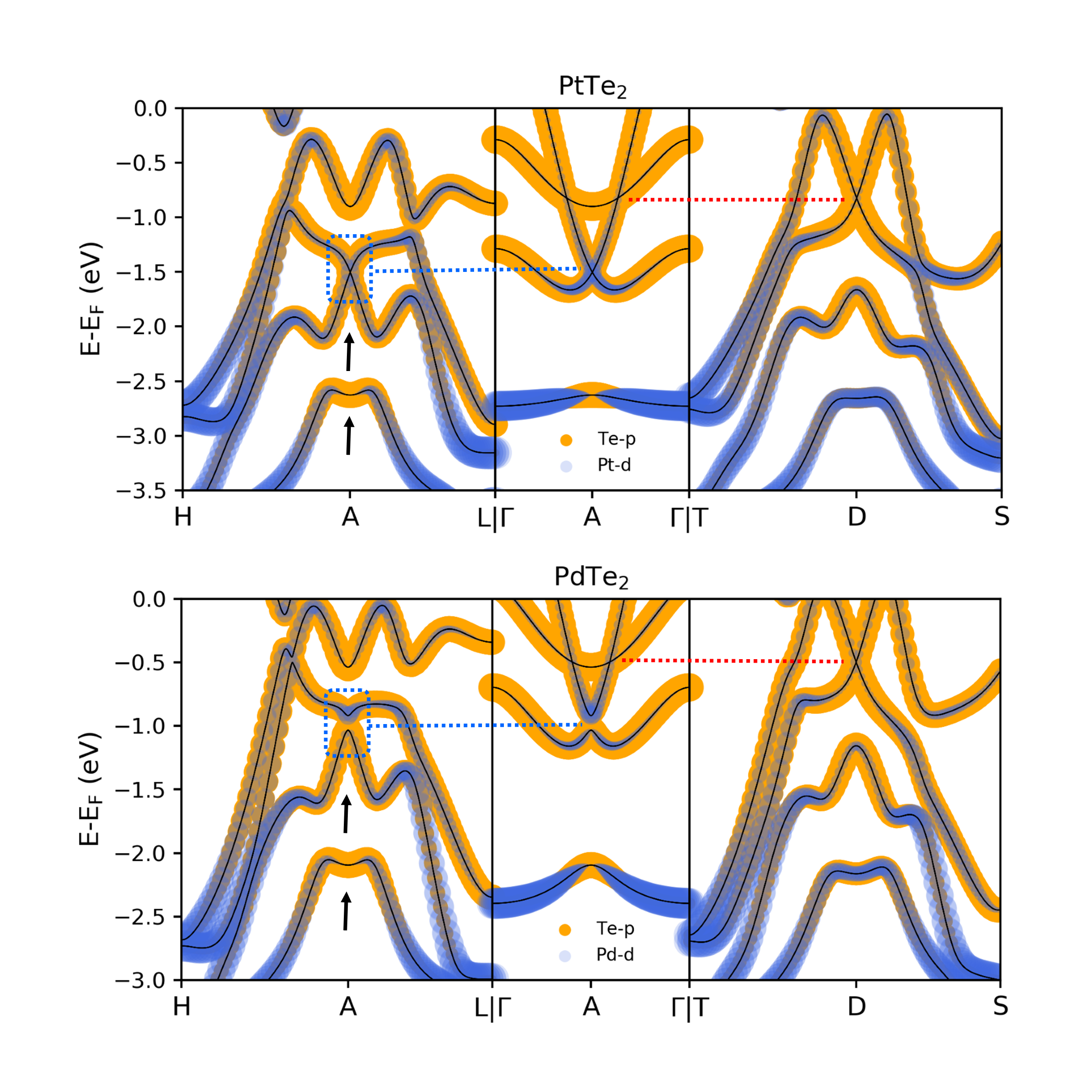}
\caption{\label{NewDFT} PtTe$_2$ and PdTe$_2$ DFT calculations showing the type-II cone along $\Gamma$A$\Gamma$ and TDS (red pointed line), the bulk bands that lead to the topologically protected surface state cone (black arrows) and the new three-dimensional dispersions (blue pointed lines). The calculations show the contribution of Te-p and Pd-d orbitals to the different bands. 
}
\end{figure}

The observed three-dimensional surface states have been identified in the past as topologically non-trivial \cite{Liu_2015,Yan2017, PhysRevB.94.121117}. Indeed, both the PtTe$_2$ and PdTe$_2$ surface states have been associated to a band inversion at TRIM A \cite{Yan2017, PhysRevB.94.121117}, and calculations of the time-reversal invariant Z$_2$ for these states have confirmed their non-trivial character \cite{Liu_2015, Yan2017, PhysRevB.94.121117}.
Our DFT calculations presented in Fig. \ref{NewDFT} are sensitive to the orbital character of the bands. They show that at TRIM $A$, where the band inversion that leads to the topologically protected surface state occurs, they are mainly of Te-p character.  
We also identify in our calculations the 3D type-II cones for both compounds, indicated by horizontal red pointed lines along $\Gamma$A$\Gamma$ and TDS. We find that bands that host the type-II excitations are also of Te-p type, as previously reported \cite{Yan2017}. Most importantly, we see evidence of the additional dispersion spotted in our ARPES data, discernable along the high symmetry directions HAL and $\Gamma$A$\Gamma$ in both PtTe$_2$ and PdTe$_2$ (indicated by pointed blue lines). Although this dispersion appears to be Dirac-like in PdTe$_2$, the point group symmetry at momentum A does not allow a 4-fold degenerate Dirac point at A. The four bands that are related to the observed dispersion form a representation where both the three-fold rotation operator and the inversion operator are diagonal \cite{TMD}. It is expected that along the $\Gamma$-A direction 
the energy dispersion is even in k$_z$ and therefore non-linear. As a consequence, this dispersion is gapped in both materials, presenting a larger gap for PdTe$_2$. Interestingly, we find that experimentally, it shows a different dichroic signal for each material, as will be developped later.  

 

\begin{figure}
\centering
\includegraphics[width=8.5cm]{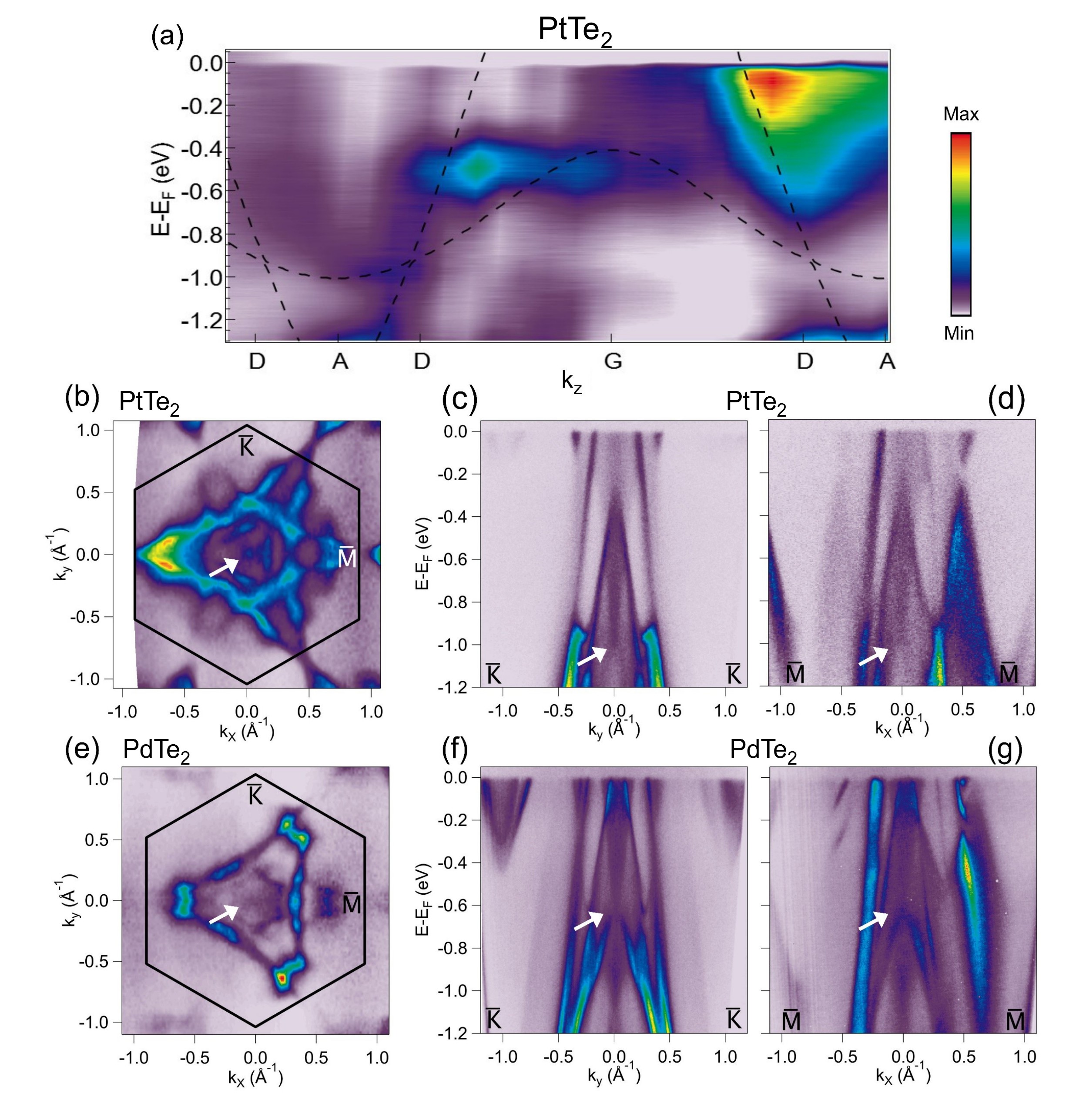}
\caption{\label{BulkConePtTe2PdTe2} 
ARPES band structure of PtTe$_2$ and PdTe$_2$ obtained using linearly p-polarized light. (a) Measured PtTe$_2$ dispersion using h$\nu$ range $\approx$ 50 - 105 eV along the k$_z$ direction at k$_x$=k$_y$=0 superimposed with DFT bulk calculations showing the type-II Dirac cone band crossing at D. (b) Data for PtTe$_2$ at h$\nu$=68 eV ($\Delta k_{z} \approx$ 0.05 c* from D) showing a constant energy map at E-E$_{\text{F}}$ $\approx$ -1 eV, corresponding to the type-II Dirac point (white arrows). (c) Dispersions along the $\overline{\text{K}}$-$\overline{\Gamma}$-$\overline{\text{K}}$ and (d) $\overline{\text{M}}$-$\overline{\Gamma}$-$\overline{\text{M}}$ directions. (e) Equivalent data for PdTe$_2$ using $h \nu$ = 60 eV ($\Delta k_{z}\approx$ 0.05 c* from D) at E-E$_{\text{F}}$ $\approx$ -0.6 eV (e) and along the $\overline{\text{K}}$-$\overline{\Gamma}$-$\overline{\text{K}}$ (f) and $\overline{\text{M}}$-$\overline{\Gamma}$-$\overline{\text{M}}$ (g) directions.
}
\end{figure}

We start by discussing the experimental signature of the linear dispersion 
that hosts exotic type-II fermions, shown in Fig. \ref{BulkConePtTe2PdTe2}.  
  Evidence of such excitations has been reported in the literature as strongly tilted Dirac cones resulting from the merging of topologically protected touching points between electron and hole pockets \cite{PhysRevB.94.121117, Yan2017}. Similar to their type-I counterparts, type-II Dirac particles are described by a massless Dirac Hamiltonian. However in the case of type-II particles, the dispersion relation is not the same in all three directions of momentum, which breaks particle-hole symmetry and the Lorentz invariance. This is prohibited in quantum field theory, as free space is isotropic, but it is permitted in a solid. The key features of such a type-II cone can be understood by adding a tilt to the Dirac Hamiltonian in the Weyl's representation: 
  $$H_{\pm}=\pm v_F\boldsymbol{\sigma}\cdot\mathbf{k}+\mathbf{t}\cdot\mathbf{k}I_2$$ 
where $v_F$ is the Fermi velocity, $\boldsymbol{\sigma}$ are the 2x2 Pauli matrices, $I_2$ the identity matrix and $\mathbf{t}$ is the tilt vector \cite{soluyanov_type-ii_2015, RevModPhys.93.025002}. In PtTe$_2$ and PdTe$_2$ the tilt is along the k$_z$ direction, where tk$_z>v_F$k$_z$.
Therefore in a type-II Dirac semimetal, instead of point-like Fermi surface at the k$_\parallel$-k$_z$ plane, there are open electron and hole pockets that touch at points. Demonstration of type-II excitations for PdTe$_2$ requires a careful comparison of the measured dispersions at different k$_z$ with first principle calculations \cite{PhysRevLett.119.016401}, as is shown in Fig. S2 of the Supp. Mat. In the case of PtTe$_2$, a clear signature of a tilted cone can be measured along the out-of-plane momentum k$_z$, as shown in Fig. \ref{BulkConePtTe2PdTe2}(a), consistent with our first principles calculations. Final state effects make the type-II cone visible only at hv=66 eV, corresponding to 
k$_z$=3.63c*. The existence of not only one but two type-II Dirac points within the same Brillouin zone (along the rotation axis of the crystal) is a consequence of the time reversal $T$ and inversion $P$ symmetries of these materials. Our DFT calculations predict this to happen at k$_z$=$\Gamma\pm$0.37c* for PtTe$_2$ and k$_z$=$\Gamma\pm$0.40c* for PdTe$_2$, that corresponds in our experiment to h$\nu$=65 eV and h$\nu$=98 eV for PtTe$_2$ and h$\nu$=62 eV and h$\nu$=99 eV for PdTe$_2$, with $\Gamma$=4c* for both compounds (check tables I and II in the Sup. Mat.). 
The type-II Dirac cones are visible along the k$_\parallel$ directions, where they appear as straight cones, seen in Figs. \ref{BulkConePtTe2PdTe2}(c), (d) for PtTe$_2$ and (f) and (g) for PdTe$_2$.  
The type-II Dirac points can also be seen in the constant energy maps shown in Figs. \ref{BulkConePtTe2PdTe2} (b) and (e), as small dots at the center of the first Brillouin zone can be discerned. Data for the pairs of type-II Dirac cones in each material are shown in Fig. S4 of the Supp. Mat.   

The use of circularly polarized light can serve as a guide to identify the three-dimensional Dirac dispersions from the topologically protected surface states. 
Despite the fact that the difference in photoemission using left- vs right- circularly polarized light known as circular dichroism (CD), is known to be strongly dependent on final states and not a reliable tool to probe the spin polarization of the initial states \cite{CDARPESBi2Te3,PhysRevB.103.245142,PhysRevB.88.241410}, CD can provide information on the initial states' relative alignment across different binding energies in the band structure \cite{PhysRevB.103.245142}. For instance, soon after the discovery of the first 3D topological insulators, CD was used to help map the spin-orbit texture of the surface states in those materials \cite{WarpedhelicalBi2Se3, wang_circular_2013}. Indeed, in the presence of spin-orbit coupling, the electron states are best described by their total angular momentum instead of their spin, and since the helicity of light in ARPES experiments couples to the electron's total angular momentum, it indirectly pairs to the spin of the material electrons \cite{schneider_spin-_1995}. 
As a result, one can measure ARPES that is helicity-dependent for linear topological surface states. 

\begin{figure}
\centering
\includegraphics[width=4cm]{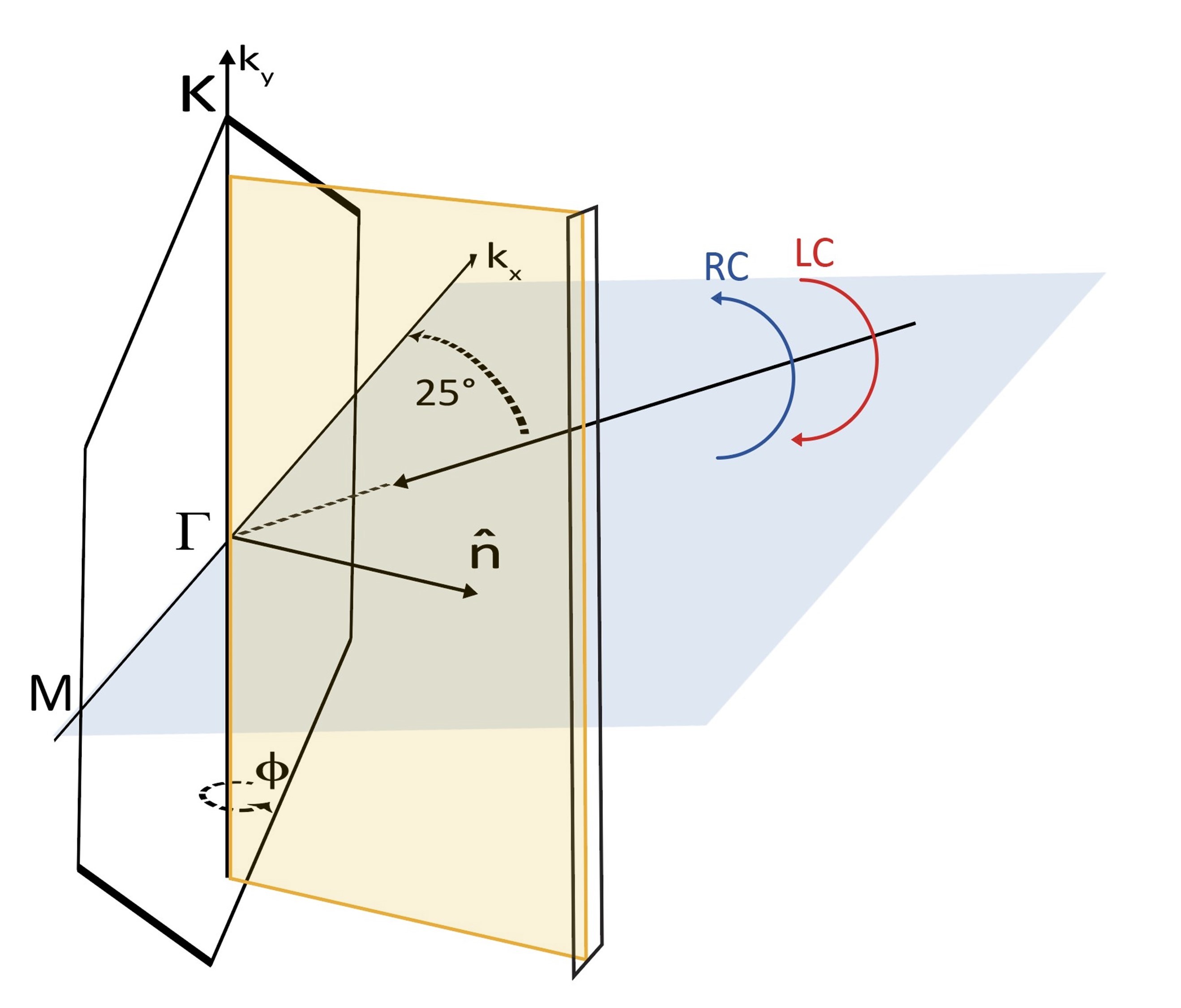}\\
\includegraphics[width=8cm]{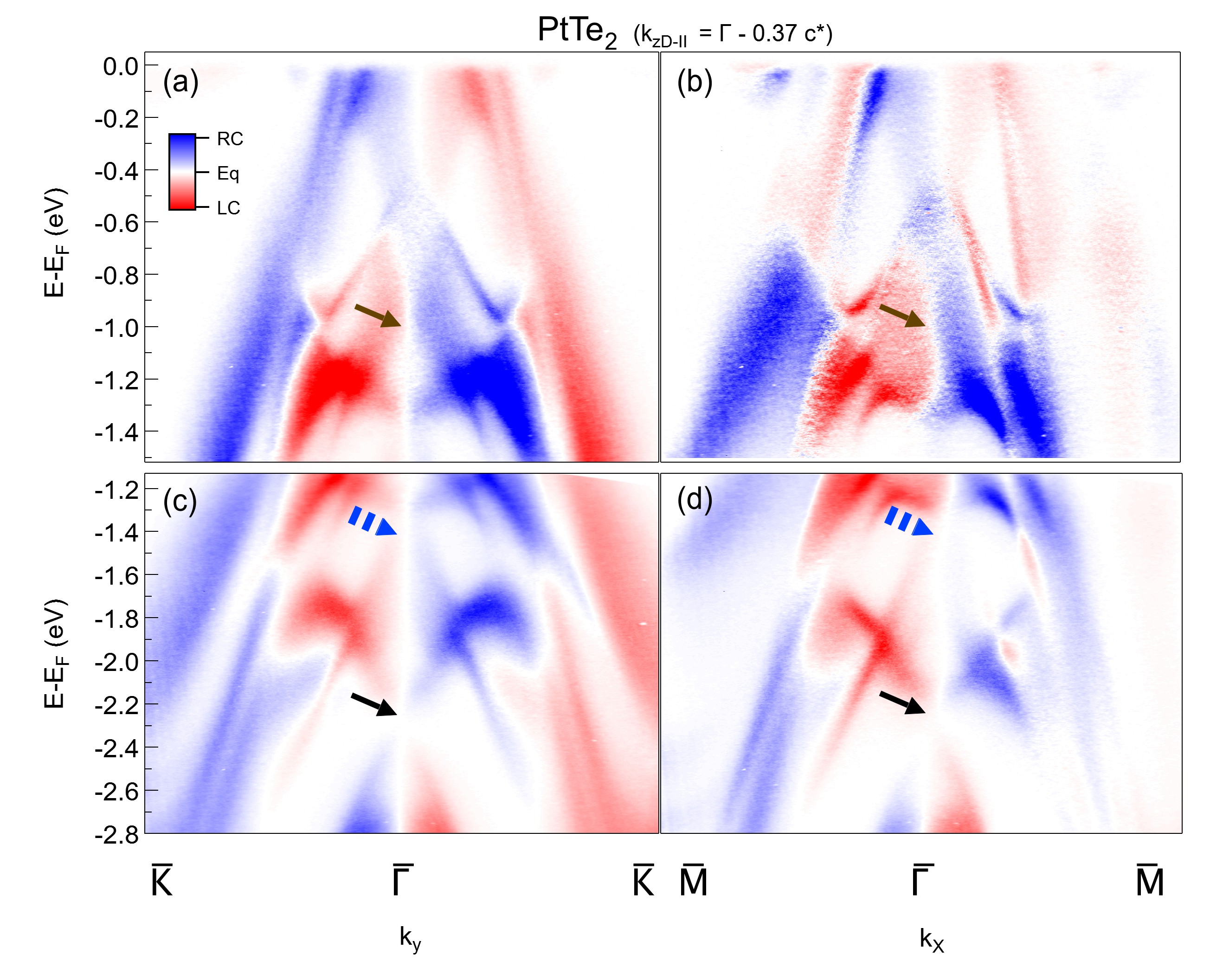}\\
\includegraphics[width=8cm]{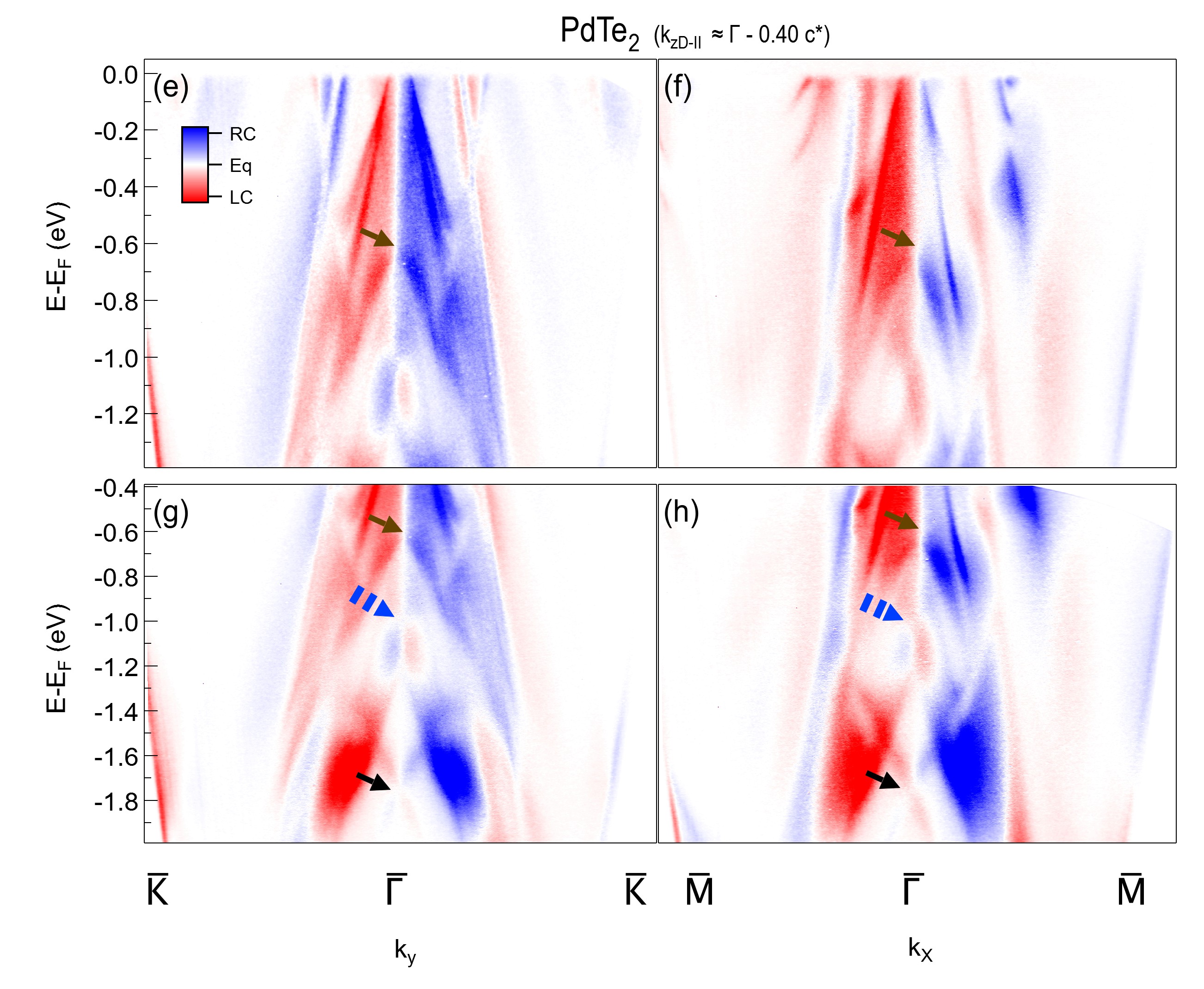}\\
\caption{\label{CircularDichroism} Top: Schematics of the experimental setup at the ALS Beamline 4. Incident photons are right-hand (blue) or left-hand (red) circularly polarized. Photons are incident
on the blue plane 25$^{\circ}$ with respect to
the surface of the sample, or 65$^{\circ}$ with
respect to the normal of the sample $\hat{n}$.
The detector plane is indicated in orange. 
$\Phi$ is the polar angle that allows to tune the angle of the sample into normal emission ($\Phi=0$). Middle: Circular polarization dependent photoemission for PtTe$_2$ and PdTe$_2$ (bottom). (a), (b) and (e), (f) show the dichroic signal for T-D-T and S-D-S respectively near the type-II Dirac cone (top brown arrows). (c), (d) and (g), (h) show the dichroic signal for the new 3D Dirac-like cones (middle blue arrows) and Dirac surface states (bottom black arrows). Data for PtTe$_2$ was taken at 66 eV photon energy, near k$_z$=3.63 c*, the type-II cone D. Data for PdTe$_2$ was taken at 60 eV, close to k$_z$=3.60c*, the D point ($\Delta k_z\approx 0.05$c* from D).}
\end{figure}

Fig. \ref{CircularDichroism} (top) shows the geometry of the experimental setup. Circularly polarized light is incident in the k$_y$=0 plane, 65$^{\circ}$ with respect to the analyzer lens, which breaks any geometrical symmetry with respect to the k$_x$=0 plane (yellow plane in Fig. \ref{CircularDichroism}). As a consequence, there is no symmetry in the photoemission measured between $\pm k_{x}$ points for different light polarizations. In contrast, there is a generic geometrical dichroism above and below the $k_y$=0 incident plane (at $\pm k_y$ for any k$_x$) provided that the crystal has also a symmetry with respect to the $k_y$=0 plane.  


We present in Fig. \ref{CircularDichroism} (a)-(h) the measured circular dichroism (CD) for PtTe$_2$ and PdTe$_2$ along the $\overline{\text{K}}$-$\overline{\Gamma}$-$\overline{\text{K}}$ and $\overline{\text{M}}$-$\overline{\Gamma}$-$\overline{\text{M}}$ directions in different binding energy ranges. We used photon energies very close to the type-II Dirac cone. To obtain a geometrical dichroism at each measurement, the azimuth angle of the sample was adjusted to have $\overline{\text{K}}$-$\overline{\Gamma}$-$\overline{\text{K}}$ or $\overline{\text{M}}$-$\overline{\Gamma}$-$\overline{\text{M}}$ along the direction of the detector (yellow plane in Fig. \ref{CircularDichroism}). As previously mentioned, the existence of a dichroic signal relies not only on the geometry of the experimental setup but also, on the symmetry of the crystal across the incident plane (blue plane in fig. \ref{CircularDichroism}). 
Because in PtTe$_2$ and PdTe$_2$ there is a mirror plane along M-$\Gamma$-A, a symmetric photoemission is observed when such mirror plane is aligned to the blue plane in Fig. \ref{CircularDichroism}, that is, along $\overline{\text{K}}$-$\overline{\Gamma}$-$\overline{\text{K}}$ rather than $\overline{\text{M}}$-$\overline{\Gamma}$-$\overline{\text{M}}$. This can be seen when comparing scans along these two high symmetry directions in Figs. \ref{BulkConePtTe2PdTe2} and \ref{CircularDichroism}. 

In general, CD is absent when a symmetry operation that transforms right into left circularly polarized light leaves the momentum of the photoelectrons \textbf{k} unaffected. This happens at the plane k$_y$=0 (blue plane in fig. \ref{CircularDichroism}), where a mirror transformation changes right into left circularly polarized light without affecting \textbf{k}, that is parallel to the normal $\hat{n}$ of the sample. There should be therefore zero CD for k$_y$=0. In contrast, along the k$_x$=0 plane (yellow plane), a reversal of the light polarization changes the sign of the y-component of \textbf{k}, resulting in a non-zero CD.  This can be observed when comparing the circular dichroism data along $\overline{\text{K}}$-$\overline{\Gamma}$-$\overline{\text{K}}$ (Figs. \ref{CircularDichroism} (a), (c), (e) and (g)) with the data along $\overline{\text{M}}$-$\overline{\Gamma}$-$\overline{\text{M}}$ (Fig. \ref{CircularDichroism} (b), (d), (f) and (h)). The former shows a clear zero CD at k$_y$=0.

Our circular dichroism measurements provide information about the distinct character of the Dirac dispersions observed in PtTe$_2$ and PdTe$_2$. Figs. \ref{CircularDichroism} (a), (b), (e) and (f) show the dichroic signal measured at binding energies close to the Fermi level. The type-II cone, expected at $\sim$ -1 eV and $\sim$ -0.6 eV  for PtTe$_2$ and PdTe$_2$ respectively (top brown arrows), show a similar dichroism for both compounds, with no switching of the dichroic signal across the Dirac point. In contrast, the surface states in these compounds previously predicted to have a non-trivial topological character (-2.25 eV for PtTe$_2$ and -1.7 eV for PdTe$_2$, bottom black arrows in Figs. \ref{CircularDichroism}(c), (d), (g) and (h)) present a clear switching of the dichroic signal across the Dirac point. This is reminiscent of the circular dichroism measurements performed during the early times of topological insulators, that were identified as mimicking the warped spin texture of the surface states in those materials \cite{WarpedhelicalBi2Se3}. The observation of a different circular dichroism for these Dirac dispersions, the type-II Dirac cone and the Dirac surface states, confirms their dissimilar nature. Although both are consequence of a band inversion, time-reversal symmetry and spin-orbit coupling, their properties differ fundamentally. As mentioned in the introduction, one is at the bulk, it is formed by a four-band crossing (the Dirac point is four-fold degenerate) and most importantly, it is only protected by crystalline symmetries. The second one is at the surface, it is non-degenerate, it has a chiral character (it is spin-polarized) and it is topologically protected. In the past, some differences in the circular dichroic signal were reported for the bulk and surface state of PdTe$_2$ at $\approx$ -0.6 eV and $\approx$ -1.7 eV respectively \cite{PhysRevLett.119.016401}. Here, we show a clear switching of the dichroic signal across the M-$\Gamma$-A plane (k$_y$=0) for the type-II cone (both in electron and hole bands), not only for PdTe$_2$ but also for PtTe$_2$ (Figs. \ref{CircularDichroism} (a), (b), and (e), (f), top brown arrows). Additionally, we see a clear difference when compared to the surface state cone in both compounds, that switches dichroic signal not only at k$_y$=0, but also at the binding energy corresponding to the Dirac point (Figs. \ref{CircularDichroism}(c), (d) and (g), (h), bottom black arrows). The fact that we observe this difference in the circular dichroism for the two Dirac dispersions in the same scan (Figs. \ref{CircularDichroism}(c), (d) and (g), (h)) rules out geometric or final state effects. 

We see in our CD data signatures of the additional dispersion already visible using linearly polarized light (Fig. \ref{83LH} blue dotted arrows), indicated also with blue arrows in Figs. \ref{CircularDichroism}(c), (d) and (g), (h). The data in Fig. \ref{CircularDichroism} were taken at a photon energy very close to the bottom type-II Dirac cone and therefore to A, where our DFT calculations predict this dispersion to appear (see Fig. \ref{NewDFT}). Interestingly, this dispersion shows no switching in the dichroic signal across the 
higher-energy and lower-energy states for PtTe$_2$ and a clear switching for PdTe$_2$. Data for another sample of PdTe$_2$, presented in Fig. S3 of the Sup. Mat., shows the same behavior. In the following, we study the dipole matrix element $M_{\mathbf{k}}=\sum_{f,in}|\langle f|\boldsymbol{A}(\omega)\cdot\boldsymbol{j}| in\rangle|^{2}$ to understand our observation, where $\boldsymbol{A}(\omega)$ is the gauge field of the incident light, $\boldsymbol{j}$ is the electron current, and the sum is over the spin-degenerate final and initial states. We consider $\boldsymbol{A}(\omega)=A_0(\omega)(\sin(\theta_0),i,-\cos(\theta_0))$ with $\theta_0=25^{\circ}$, the incident light angle with respect to the surface of the sample (see Fig. \ref{CircularDichroism}). The initial state $\ket{in}$ is a Bloch state generally expressed as $\ket{in}=\sum_{l}u_{l\mathbf{k}}\ket{l_{in}}$ with $\ket{l_{in}}, and \ l=1,2,3,4$ corresponds to the four eigenstates of the $C_{3}$ operator. For the final states, we consider spin-degenerate states in the bulk, denoting them as $\ket{\uparrow_{f}}$ and $\ket{\downarrow_{f}}$. By considering the constraints imposed by time-reversal symmetry, inversion symmetry, three-fold rotation symmetry, and the mirror-y symmetry, we identify that only $\bra{\uparrow_{f}}j_{z}\ket{1/3_{i}}$, $\bra{\downarrow_{f}}j_{z}\ket{2/4_{i}}$, $\bra{\uparrow_{f}}j_{x}-i j_{y}\ket{2/4_{i}}$, $\bra{\downarrow_{f}}j_{x}+i j_{y}\ket{1/3_{i}}$ are nonzero, which allows us to write down a close formula for $M_{\mathbf{k}}$. We then calculate the difference of $M_{\mathbf{k}}$ for the right and left circularly polarized light, $\Delta M_{\mathbf{k}}$, and reveal its dependence on $u_{l\mathbf{k}}$. Details of the symmetry analysis and calculation of $\Delta M_{\mathbf{k}}$ can be found in the Supplementary Materials.

To understand the different dichroism observed in PtTe$_{2}$ and PdTe$_{2}$, we consider a symmetry allowed $\mathbf{k}\cdot\mathbf{p}$ Hamiltonian around the $A$ point for the four related bands:
\begin{equation}
\label{eq:H}
\begin{split}   
H(k_{x},k_{y},k_{z}) & =t_{1}(k_{x}\tau_{x}\sigma_{x}+k_{y}\tau_{x}\sigma_{y})+(M+\lambda k_{z}^2)\tau_{z}\sigma_{0} \\
& +t_{2}i(k_{+}^{3}-k_{-}^{3})\tau_{x}\sigma_{z},
\end{split}
\end{equation}
where $t_1$ is a parameter of the form $\hbar v$ with $v$ the group velocity of the modes near the high symmetry point $A$, and $t_2$ has units of energy/lenght$^3$. $M$ is a parameter that quantifies the energy separation of the four bands at $A$, $\lambda$ is of the form $\hbar^2/2m_{eff}$, with $m_{eff}$ the effective mass of the electrons near the high symmetry point $A$ and $k_{\pm}=k_x\pm i k_y$. The introduction of terms linear in $k_{x}$ and $k_{y}$ is natural and finds support from our ARPES data and DFT calculations. The term cubic in $k_{x}$ and $k_{y}$ is symmetry allowed and relevant to the dichroism. If the linear term dominates ($t_{1}\gg t_{2}$), the eigenstates lead to a $\Delta M_{f\bold{k}}$ that has the same sign for the higher and lower energy bands, but changes sign for opposite momentum, consistent with our observation for PtTe$_2$. On the other hand, if the cubic term in $k$ dominates, $\Delta M_{f\bold{k}}$ will change sign upon both energy flip and momentum flip, which is consistent with the observation for PdTe$_2$. We propose that despite PtTe$_2$ and PdTe$_2$ belonging to the same space group, a cubic term can be significant in one material but negligible in the other. This difference leads to discernible dichroism behaviors between the two materials. This understanding is also consistent with our DFT calculations in Fig.~\ref{NewDFT}, where the dispersion of PtTe$_2$ (black arrows) appears to be more linear. 

The symmetry operations allowed by the crystallographic space group P$\bar{3}$m1 are at the origin of the most important electronic dispersions in PtTe$_2$ and PdTe$_2$, namely, the bulk type-II Dirac cones and the topologically protected surface states. Indeed, it has been demonstrated that a perturbation that breaks the trigonal symmetry of the system not only removes the type-II Dirac cones, but also significantly affects the surface states, which are observable in ARPES experiments \cite{PhysRevB.94.121117}. Here we have shown that even symmetry preserved perturbations can lead to experimentally discernible ARPES features by using circularly polarized light.  

\begin{figure}
\centering
\includegraphics[width=8.5cm]{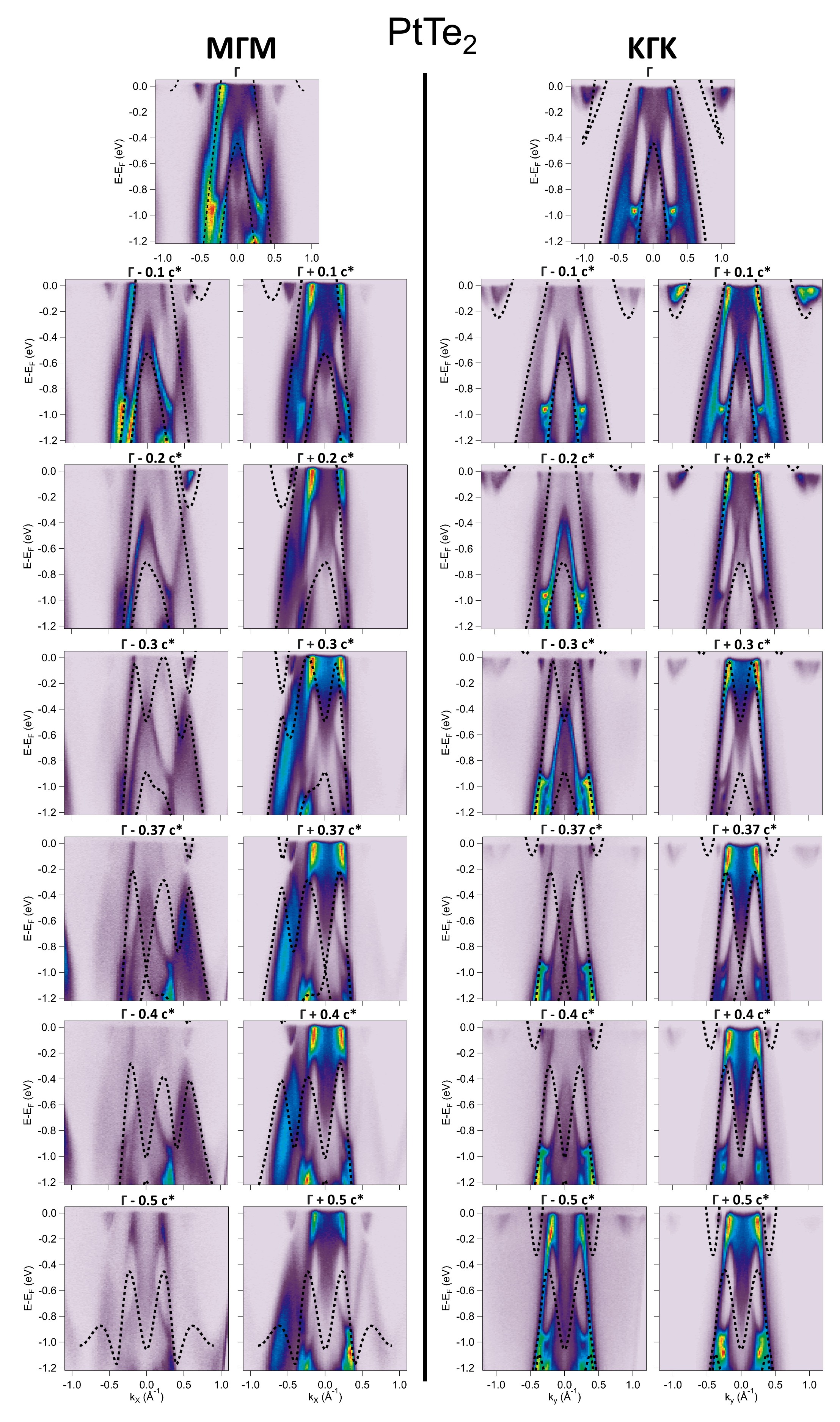}
\caption{\label{MGMKGK} Left: PtTe$_2$ dispersions along the $\overline{\text{M}}$-$\overline{\Gamma}$-$\overline{\text{M}}$ direction at k$_z$ values ranging from $\Gamma-0.5$c* to $\Gamma+0.5$c*, corresponding to the high symmetry point A ($\Gamma$ = 4c*). Data puts in evidence the inversion symmetry of the crystal, where E$_{\sigma}$(k$_x$, k$_y$=0, k$_z$)=E$_{\sigma}$(-k$_x$, k$_y$=0, -k$_z$), see text. Right: PtTe$_2$ dispersions along the $\overline{\text{K}}$-$\overline{\Gamma}$-$\overline{\text{K}}$ direction at the same k$_z$ values. k$_z$=$\Gamma\pm$0.37 c* shows the formation of the type-II cone. Data was taken using p-polarized light with h$\nu$ range $\approx$ 60 - 105 eV.}

\end{figure}

Finally, it is worth noting that overall, some of the symmetries of the crystals PtTe$_2$ and PdTe$_2$ can be easily identified in the measured and calculated band structure for these materials. For example, it can be noted that the dispersion along the $\overline{\text{K}}$-$\overline{\Gamma}$-$\overline{\text{K}}$ direction is always symmetric with respect to $\Gamma$ in both compounds (Figs. \ref{BulkConePtTe2PdTe2}(c) and (f)), in contrast to the dispersion along the $\overline{\text{M}}$-$\overline{\Gamma}$-$\overline{\text{M}}$ direction (Figs. \ref{BulkConePtTe2PdTe2}(d) and (g)). This three-fold symmetry is consequence of the centrosymmetric trigonal crystal structure of the PtTe$_2$ and PdTe$_2$ crystals, with mirror planes along M-$\Gamma$-A as previously described. 
Additionally, the inversion symmetry present in these crystals relates the energy of the bands at opposite momenta, E$_{\sigma}$(k$_x$, k$_y$, k$_z$)=E$_{\sigma}$(-k$_x$, -k$_y$, -k$_z$), 
seen in our data and DFT calculations. Fig. \ref{MGMKGK} shows the measured dispersions for PtTe$_2$ superimposed to our DFT calculations along $\overline{\text{M}}$-$\overline{\Gamma}$-$\overline{\text{M}}$ and $\overline{\text{K}}$-$\overline{\Gamma}$-$\overline{\text{K}}$. We compare dispersions at opposite k$_z$ with respect to k$_z$=4c$^*$ ($\Gamma$, h$\nu$=81 eV), in a range from k$_z$=$\Gamma$-0.5c$^*$ (A, h$\nu$=60 eV) to k$_z$=$\Gamma$+0.5c$^*$ (A, h$\nu$=104 eV). 
In that data, we see a clear signature of the inversion symmetry. As expected, the data is symmetric with respect to k$_y$=0 \AA $^{-1}$ along the $\overline{\text{K}}$-$\overline{\Gamma}$-$\overline{\text{K}}$ direction for all k$_z$, consistent with the fact that a mirror plan lies along the M-$\Gamma$-A or k$_y$=0 \AA $^{-1}$ plane. In contrast, along the $\overline{\text{M}}$-$\overline{\Gamma}$-$\overline{\text{M}}$ direction, the data is only symmetric with respect to the M-$\Gamma$-A or k$_x$=0 \AA$^{-1}$ plane at the high symmetry points k$_z$=$\Gamma$ and k$_z$=A as expected (left top and bottom figures). For other k$_z$ momenta, we can see in our data and DFT calculations that E$_{\sigma}$(k$_x$, k$_y$=0, k$_z$)=E$_{\sigma}$(-k$_x$, k$_y$=0, -k$_z$), consistent with inversion symmetry. The discussed type-II cone in PtTe$_2$ is visible at k$_z$=$\Gamma\pm$0.37c*.

\section{Conclusions}

In conclusion, we have performed a detailed study of the electronic structure of the type-II Dirac semimetals PtTe$_2$ and PdTe$_2$ using high resolution ARPES and DFT calculations, unveiling a new three-dimensional dispersion in these compounds, at -1.4 eV for PtTe$_2$ and -1.1 eV for PdTe$_2$. We observe a distinct signature of such dispersion in PtTe$_2$ and PdTe$_2$ when using circularly polarized light. While PdTe$_2$ shows a clear switching of the dichroic signal across the higher-energy and lower-energy states, the switching is absent for PtTe$_2$. We relate this observation to the symmetry of the dipole matrix element. After considering the most general Hamiltonian that complies with all the symmetries of both PtTe$_2$ and PdTe$_2$, the contrasting CD signal for both compounds can be explained if one considers that a linear term in momentum dominates in the case of PtTe$_2$ and a cubic term in the case of PdTe$_2$. It is interesting to realize that while the two cases may be hard to distinguish by just looking at the dispersion, 
there is a clear difference when analyzing the CD.   

Our CD measurements put also in evidence the prominent difference between the previously reported Dirac dispersions in these materials. 
The type-II Dirac cone that lives in the bulk, formed by four bands, protected by the symmetries of the crystal 
and hosting exotic type-II Dirac fermions, shows no switching of the dichroic signal across the Dirac point. In contrast, the surface state Dirac cone that is formed at the bandgap between two bands, topologically protected and hosting chiral Dirac fermions, shows a clear switching of the dichroic signal across the Dirac point. 

Our work not only provides a useful reference for the characterization of other transition metal dichalcogenides with topological properties, but more generally, it shows how circular dichroism in ARPES can be useful in differentiating the topological character of two otherwise equivalent band dispersions, and to identify the different attributes in the band structure of similar materials.

\section{Acknowledgements}
The primary funding for this work was provided by the U.S. Department of Energy, Office of Science, Office of Basic Energy Sciences under Contract No. DE-SC0018154. I.P, D.B and M.N. were supported by the Cal State Long Beach and Ohio State University Partnership for Education and Research in Hard and Soft Materials, a National Science Foundation PREM, under Grant No. 2122199 for traveling to the ALS. JEG, AJW, and WLBH acknowledge the Center for Emergent Materials, an NSF MRSEC, under award number DMR-2011876 for crystal growth. JQ and WW acknowledge the Air Force Office of Scientific Research for funding from grant number FA9550-21-1-02684 for theoretical calculations. Calculations were performed at the Ohio Supercomputer Center under project number PAS0072. The Advanced Light Source is supported by the Director, Office of Science, Office of Basic Energy Sciences, of the U.S. Department of Energy (U.S. DOE-BES) under Contract No. DE-AC02-05CH11231. The photoemission work by L. M. and A. L. was supported by US Department of Energy, Office of Science, Office of Basic Energy Sciences, Materials Sciences and Engineering Division under contract number DE-AC02- 05CH11231 within the vdW heterostructure Program (KCWF16). COA would like to acknowledge useful discussions with Jean Noel Fuchs, Frederique Piechon and Mark Goerbig. I.P., D.B, M.N. would like to acknowledge internal CSULB fellowships such as the John E. Frederickson Scholarship, the graduate RSCA award, the graduate travel fellowship, the graduate student honors award, the Richard D. Green Graduate Research Fellowship from the College of Natural Sciences and Mathematics at CSULB and the Google American Physical Society Inclusive Graduate Education Network Bridge Fellowship Program at CSULB.

\section*{Competing interests}
The authors declare no competing interests.

\bibliography{BibliographyPtTe2}

\onecolumngrid
\section*{Supplementary Materials}
\renewcommand{\thefigure}{S\arabic{figure}} 

\subsection{Unit cell of PtTe$_2$ and PdTe$_2$ space group}

\begin{figure}[H]
\begin{center} 
\includegraphics[width=4cm]{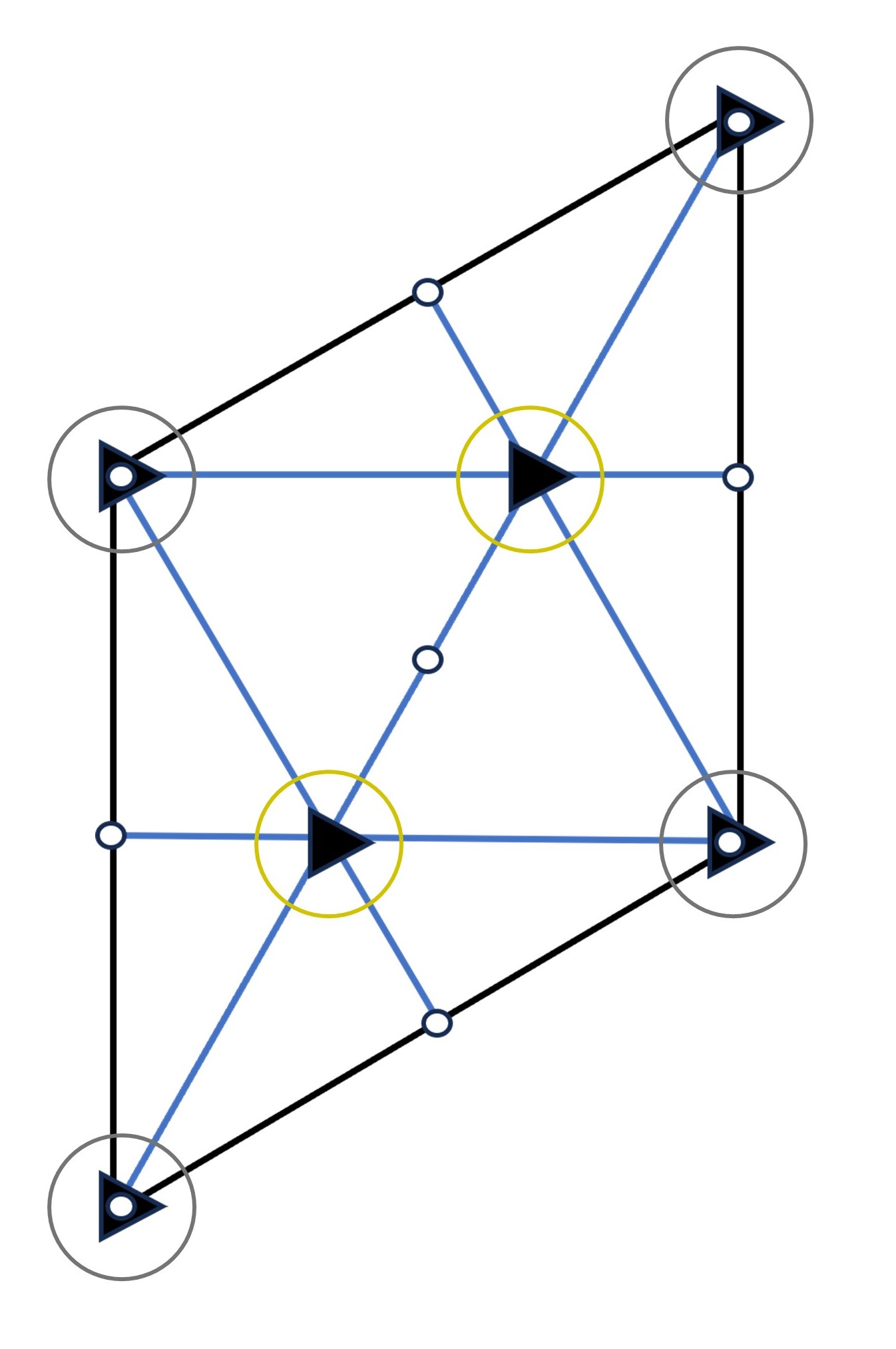}
\end{center}
\caption{\label{cell}Real space in-plane unit cell of space group 164 adapted from Volume A of ref. \cite{2016}. The blue lines represent the mirror planes of the crystal and the triangles represent centers of three-fold rotation (C$_3$). The white circles represent inversion centers, and the black triangles with white circles correspond to linear combinations of the in-plane unit cell lattice vectors. The transition metal atoms (Pt or Pd) and Te atoms are represented with gray and gold circles, respectively.}
\end{figure}

Figure \ref{cell} shows the unit cell of space group P$\bar{3}$m1 (No 164) associated to PtTe$_2$ and PdTe$_2$. Bold black lines represent the three mirror planes of the lattice, triangles the threefold rotation centers and white circles the inversion centers.

\subsection{PdTe$_2$: Characterization of the band structure at different photon energies}
\begin{figure}[H]
\includegraphics[width=16cm]{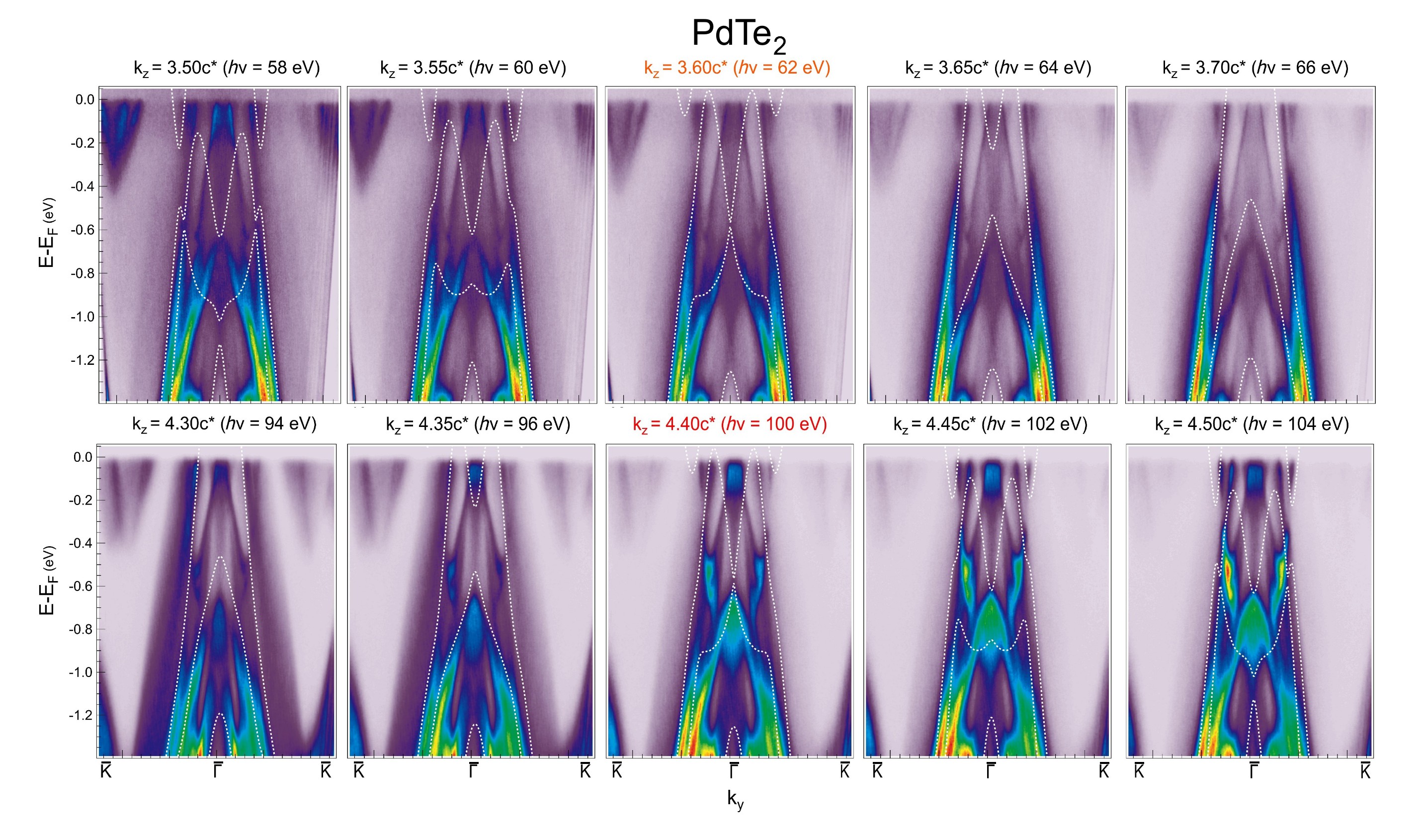}
\caption{ARPES data and DFT calculations for PdTe$_2$ represented in a photon energy range corresponding to k$_z=\Gamma\pm$ 0.5c*, with $\Gamma=4c*$ showing the formation of two type-II Dirac cones on the k$_z$ axis, at k$_z=3.60c*$ and k$_z=4.40c*$, within the same Brillouin zone.\label{PdTe2kz_Mahfuzun}}
\end{figure}

Fig. \ref{PdTe2kz_Mahfuzun} shows the band structure for PdTe$_2$ for different photon energies, superimposed to our DFT calculations. We represent the data in the k$_z$ range corresponding to A-D-$\Gamma$-D-A. It can be noted the formation of a type-II Dirac cone at k$_z$=3.6c* and k$_z$=4.4c* that is, k$_z=\Gamma\pm$ 0.4c*. As the Dirac cone is tilted, it is natural that is doesn't appear at k$_z$=$\Gamma$, but at opposite locations on the k$_z$ axis, in compliance with the time reversal $T$ and inversion $P$ symmetries of the crystal. 

\subsection{Circular dichroism of the new three-dimensional Dirac-like dispersion}
\begin{figure}[H]
\includegraphics[width=5cm]{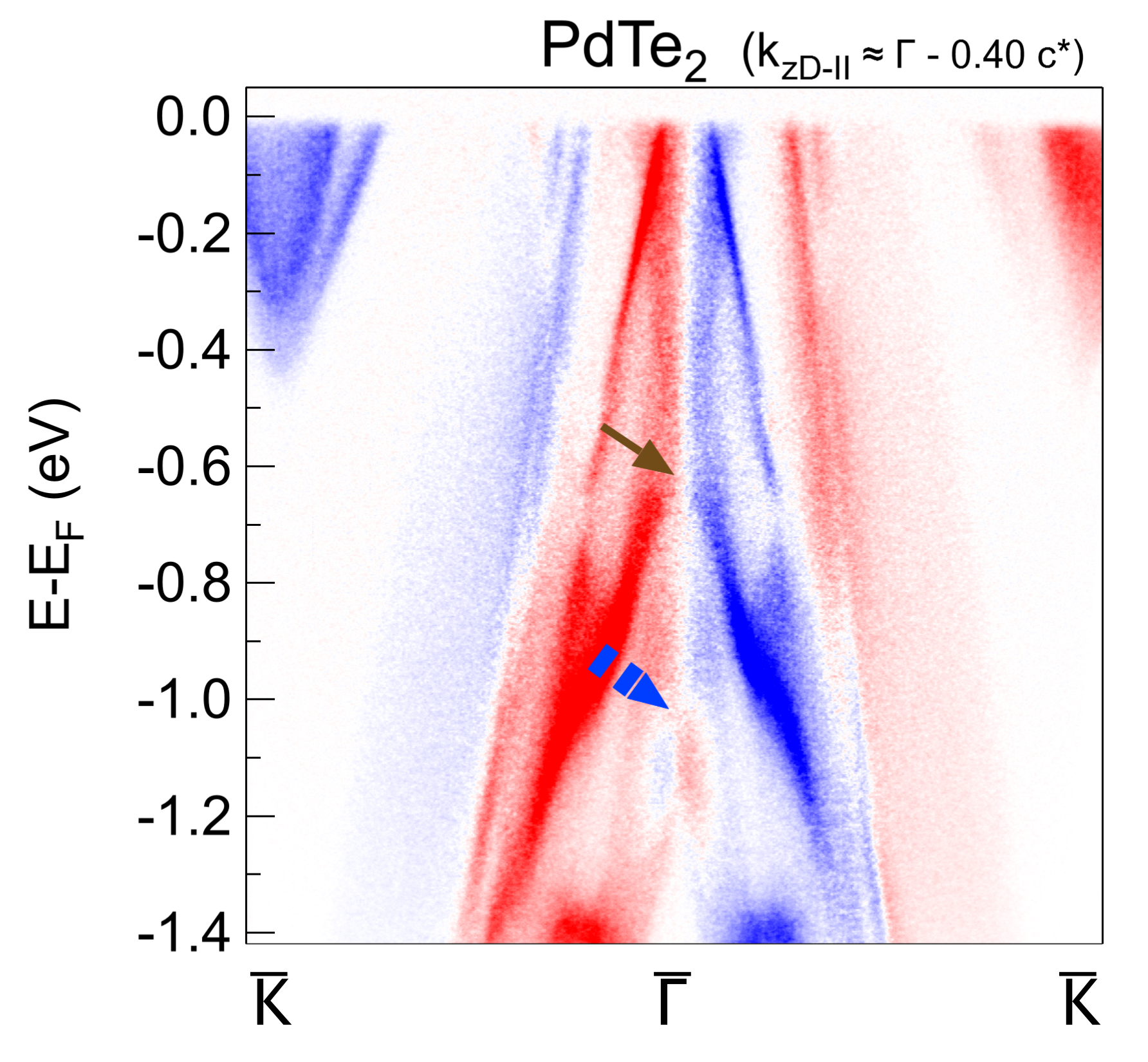}
\caption{\label{PdTe2AdditionalCD}Circular dichroism for a second sample of PdTe$_2$ taken at h$\nu$=60 eV (k$_z$=3.5c*) along the $\overline{\text{K}}$-$\overline{\Gamma}$-$\overline{\text{K}}$ direction, close to T-D-T ($\Delta k_z\approx 0.1$c* from D). The binding energy of the type-II cone is shown by the brown arrow and the dichroic signal for the new 3D Dirac-like dispersion is indicated by the blue arrow.}
\end{figure}

Fig. \ref{PdTe2AdditionalCD} shows the circular dichroism of a different sample of PdTe$2$ taken close to the type-II Dirac cone, showing a similar dichroic signal as shown in the main text.

\subsection{Pair of type-II Dirac cones for PtTe$_2$ and PdTe$_2$}

\begin{figure}[H]
\centering
\includegraphics[width=9cm]{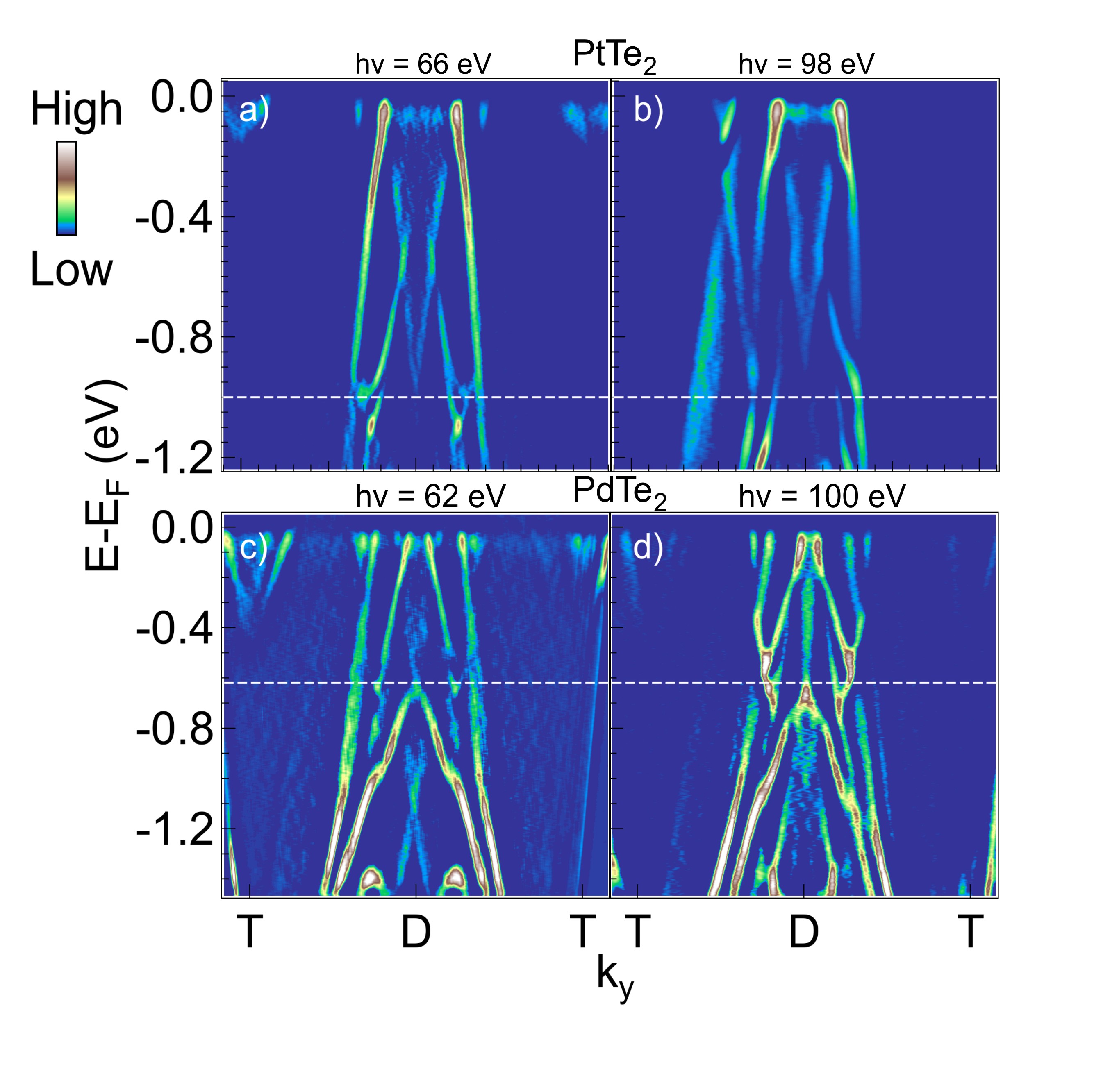}
\caption{\label{PairConesPtTe2PdTe2} Pair of Type-II Dirac cones in PtTe$_2$ and PdTe$_2$. Top: dispersions for PtTe$_2$ showing the type-II cones at 66 eV (left) and 98 eV (right) corresponding to $k_{z}=3.63$ c* and $k_{z}=4.37$ c*
respectively. Bottom: equivalent dispersions for PdTe$_2$ at 62 eV (left) and 100 eV (right) corresponding to $k_{z}=3.60$ c* and $k_{z}=4.40$ c*
respectively. White pointed lines indicate the Dirac points.}
\end{figure}

\begin{table}[h!]
    \centering
    \caption{PtTe$_{2}$ k$_{z}$-points and corresponding photon energies calculated inner potential 
    V$_0$=12 eV and out-of-plane lattice constant c=5.224 \AA. The Fermi level binding energy was used for points A and $\Gamma$. The type-II Dirac point binding energy of 1.0 eV was used for points D.}
    \begin{tabular}{ccc}
        \hline & & \\
        k-point & k$_{z}$ (c*) & h$\nu$ (eV)\\
        \hline & & \\
        A & 3.50 & 60.0 \\
        D & 3.63 & 65.2 \\
        $\Gamma$ & 4.00 & 80.7 \\
        D & 4.37 & 97.7 \\
        A & 4.50 & 104 \\
        \hline
    \end{tabular}
\end{table}

\begin{table}[h!]
    \centering
    \caption{PdTe$_{2}$ k$_{z}$-points and corresponding photon energies calculated using inner potential V$_0$=16$\pm$1 eV and out-of-plane lattice constant c=5.130 \AA. The Fermi level binding energy was used for points A and $\Gamma$. The type-II Dirac point binding energy of 0.6 eV was used for points D.}
    \begin{tabular}{ccc}
        \hline & & \\
        k-point & k$_{z}$ (c*) & h$\nu$ (eV)\\
        \hline & & \\
        A & 3.50 & 58.2 \\
        D & 3.60 & 62.3 \\
        $\Gamma$ & 4.00 & 79.6 \\
        D & 4.40 & 98.9 \\
        A & 4.50 & 104 \\
        \hline
    \end{tabular}
\end{table}

\subsection{$\mathbf{k}\cdot\mathbf{p}$ Hamiltonian around high symmetry point A for the Dirac-like dispersion}
Here, we elaborate that Eq.~(1) in the main text is a symmetry allowed Hamiltonian. We choose the basis where both $H(0,0,k_{z})$ and $C_{3z}$ are diagonal, under which the Hamiltonian along the rotation axis can be generally expressed as:
\begin{equation}
\label{eq:Hkz}
H(0,0,k_{z})=d_{0}+d_{1}\tau_{3}\sigma_{0}+d_{2}\tau_{3}\sigma_{3}+d_{3}\tau_{0}\sigma_{3},
\end{equation}
where $\tau$'s and $\sigma$'s are Pauli the matrices for orbital and spin, respectively. Considering the time reversal operator $T=i\sigma_{y}K$, and the inversion operator $I=-\tau_{z}\sigma_{0}$ that are consistent with Ref.~\onlinecite{TMD} we obtain that only the $d_{0}$ and $d_{1}$ terms in Eq.~\eqref{eq:Hkz} are compatible with time-reversal and inversion symmetry, and thus
\begin{equation}
\label{eq:Hkz1}
H(0,0,k_{z})=d_{0}+d_{1}(k_{z})\tau_{3}\sigma_{0}, \ d_{1}(k_{z})=d_{1}(-k_{z}).
\end{equation}
This indicate that the four bands around the high symmetry point $A$ cannot have a linear dispersion along the $k_z$-direction. Away from the rotation axis, we consider terms linear and cubic in $k_{x}$ and $k_{y}$ in Eq.~(1) in the main text. Note that we have not written down all symmetry allowed terms, we only write down two terms of which the orders in $k_{x}$ and $k_{y}$ are low. Interestingly, only with these two terms, we can already explain the different dichroism behaviors observed in PtTe$_2$ and PdTe$_2$. We leave a comprehensive analysis of all symmetry allowed terms into future research.

\subsection{Symmetry analysis of the dipole matrix element and the dichroism}
Here, we first conduct a symmetry analysis for dipole matrix element $\bra{\uparrow/\downarrow_{f}}\mathbf{j}\ket{l_{i}}$. Then, we derive a close formula for $M_{\mathbf{k}}=\sum_{f,in}|\bra{f}\mathbf{A}\cdot \mathbf{j}\ket{in}|^2$. Finally, we write down the difference of $M_{\mathbf{k}}$ for right and left circular light, $\Delta M_{\mathbf{k}}$, which directly captures the dichroism.  

Under the basis we use, the symmetry operators have the following representations: Time reversal $T=i\tau_{0}\sigma_{2}K$; inversion $I=\tau_{3}\sigma_{0}$; three-fold rotation $C_{3}=\operatorname{diag}\{e^{-i\pi/3},e^{i\pi/3},e^{-i\pi/3},e^{i\pi/3}\}$; and the mirror-y $m_{y}=i\tau_{0}\sigma_{x}$. With these setups, we now consider the constraints on matrix elements of $j_{z}$ and $j_{\pm}=j_{x}\pm ij_{y}$ imposed by these symmetries.  First, from trivial equality $\bra{f}C_{3}^{-1}C_{3}j_{z/\pm}C_{3}^{-1}C_{3}\ket{in}=\bra{f}j_{z/\pm}\ket{in}$ and the angular momentum carried by $\ket{f}$ and $\ket{in}$, we can derive that (i)$\bra{\uparrow_{f}}j_{z}\ket{l_{in}}$ ($\bra{\downarrow_{f}}j_{z}\ket{l_{in}}$) can be nonzero only when $l=1,3$ ($l=2,4$); (2)$\bra{\uparrow_{f}}j_{-}\ket{l_{in}}$ ($\bra{\downarrow_{f}}j_{+}\ket{l_{in}}$) can be nonzero only when $l=2,4$ ($l=1,3$); and (3) all other matrix elements are zero. Next, the time-reversal symmetry relates states with opposite spins, leading to constraints: 
$\bra{\uparrow_{f}}j_z\ket{l_{in}}=-(\bra{\downarrow_{f}}j_{z}\ket{(l+1)_{i}})^{\star}$ and $\bra{\downarrow_{f}}j_{+}\ket{l_{in}}=(\bra{\uparrow_{f}}j_{-}\ket{(l+1)_{i}})^{\star}$ for $l=1,3$. Finally, the mirror-y symmetry ensures that $\bra{\uparrow_{f}}j_{z}\ket{l_{in}}=\bra{\downarrow_{f}}j_{z}\ket{(l+1)_{i}}$ and $\bra{\downarrow_{f}}j_{+}\ket{l_{in}}=\bra{\uparrow_{f}}j_{-}\ket{(l+1)_{i}}$ for $l=1,3$. It's important to note that we assume the final state to be a superposition of inversion-parity even and odd states (e.g., a plane wave), hence the inversion symmetry does not impose any constraints on the matrix elements. Collecting all these constraints, we can conclude that $\bra{\uparrow_{f}}j_{z}\ket{1/3_{i}}=\bra{\downarrow_{f}}j_{z}\ket{2/4_{i}}$ are purely imaginary; $\bra{\uparrow_{f}}j_{x}-i j_{y}\ket{2/4_{i}}=\bra{\downarrow_{f}}j_{x}+i j_{y}\ket{1/3_{i}}$ are purely real; and all other matrix elements are vanishing. For simplicity in the following elaborations, we set
\begin{equation}
\label{eq:matrixele}
\begin{aligned}
&\bra{\uparrow_{f}}j_{z}\ket{1_{in}}=\bra{\downarrow_{f}}j_{z}\ket{2_{in}}=i\alpha_{1}, \ \bra{\uparrow_{f}}j_{z}\ket{3_{in}}=\bra{\downarrow_{f}}j_{z}\ket{4_{in}}=i\alpha_{3};
\\
&\bra{\uparrow_{f}}j_{x}\ket{2_{in}}=\bra{\downarrow_{f}}j_{x}\ket{1_{in}}=\alpha_{2},\ \bra{\uparrow_{f}}j_{x}\ket{4_{in}}=\bra{\downarrow_{f}}j_{x}\ket{3_{in}}=\alpha_{4};
\\
&\bra{\uparrow_{f}}j_{y}\ket{2_{in}}=-\bra{\downarrow_{f}}j_{y}\ket{1_{in}}=i\alpha_{2}, \ \bra{\uparrow_{f}}j_{y}\ket{4_{in}}=-\bra{\downarrow_{f}}j_{y}\ket{3_{in}}=i\alpha_{4}.
\end{aligned}
\end{equation}
Then, $M_{\mathbf{k}}=\sum_{f,in}|\bra{f}\mathbf{A}\cdot \mathbf{j}\ket{in}|^2$ can be expressed in terms of these real parameters and the Bloch states $u_{l\mathbf{k}}$:
\begin{equation}
\begin{aligned}
M_{f\mathbf{k}}&=\sum_{in}|i(\alpha_{1}u_{1\mathbf{k}}+\alpha_{3}u_{3\mathbf{k}})A_{z}+(\alpha_{2}u_{2\mathbf{k}}+\alpha_{4}u_{4\mathbf{k}})(A_{x}+i A_{y})|^2
 \\
&+|i(\alpha_{1}u_{2\mathbf{k}}+\alpha_{3}u_{4\mathbf{k}})A_{z}+(\alpha_{2}u_{1\mathbf{k}}+\alpha_{4}u_{3\mathbf{k}})(A_{x}-i A_{y})|^2,
\end{aligned}
\end{equation}
where the $\sum_{in}$ is the sum over spin-degenerate initial states.
Flipping the $A_{y}$ component of the vector potential or flipping both $A_{x}$ and $A_{z}$ components can transform right circular light into left circular light. For simplicity, in the following we consider $A_{y}\rightarrow -A_{y}$ to calculate the difference in $M_{\mathbf{k}}$ for the right and left circular light:

\begin{equation}
\label{eq:difference}
\begin{aligned}
\Delta M_{f\mathbf{k}}&=\sum_{in}4\operatorname{Im}(A_{x}A_{y}^{\star})(|\alpha_{2}u_{2\mathbf{k}}+\alpha_{4}u_{4\mathbf{k}}|^2-|\alpha_{2}u_{1\mathbf{k}}+\alpha_{4}u_{3\mathbf{k}}|^2)
\\
&+4\operatorname{Re}\{A_{z}A_{y}^{\star}[(\alpha_{1}u_{1\mathbf{k}}+\alpha_{3}u_{3\mathbf{k}})(\alpha_{2}u_{2\mathbf{k}}+\alpha_{4}u_{4\mathbf{k}})^{\star}-(\alpha_{1}u_{2\mathbf{k}}+\alpha_{3}u_{4\mathbf{k}})(\alpha_{2}u_{1\mathbf{k}}+\alpha_{4}u_{3\mathbf{k}})^{\star}]\},
\end{aligned}
\end{equation}
where 
\begin{equation}
\label{eq:difference1}
\begin{aligned}
\sum_{in}|\alpha_{2}u_{2\mathbf{k}}+\alpha_{4}u_{4\mathbf{k}}|^2-|\alpha_{2}u_{1\mathbf{k}}+\alpha_{4}u_{3\mathbf{k}}|^2&=\sum_{in}\alpha_{2}^2(|u_{2\mathbf{k}}|^2-|u_{1\mathbf{k}}|^2)+\alpha_{4}^2(|u_{4\mathbf{k}}|^2-|u_{3\mathbf{k}}|^2)
+2\alpha_{2}\alpha_{4}\operatorname{Re}(u_{2\mathbf{k}}^{\star}u_{4\mathbf{k}}-u_{1\mathbf{k}}^{\star}u_{3\mathbf{k}})
\\
&=\sum_{in}2\alpha_{2}\alpha_{4}\operatorname{Re}(u_{2\mathbf{k}}^{\star}u_{4\mathbf{k}}-u_{1\mathbf{k}}^{\star}u_{3\mathbf{k}}).
\end{aligned}
\end{equation}
The last step is because $\sum_{in}(|u_{2\mathbf{k}}|^2-|u_{1\mathbf{k}}|^2)$ and $\sum_{in}(|u_{4\mathbf{k}}|^2-|u_{3\mathbf{k}}|^2)$ are basically $-\langle\sigma_{z}\rangle$ in different sectors, which are forced to be zero by time-reversal and inversion symmetries. Similarly, we have
\begin{equation}
\label{eq:difference2}
\begin{aligned}
&\sum_{in}(\alpha_{1}u_{1\mathbf{k}}+\alpha_{3}u_{3\mathbf{k}})(\alpha_{2}u_{2\mathbf{k}}+\alpha_{4}u_{4\mathbf{k}})^{\star}-(\alpha_{1}u_{2\mathbf{k}}+\alpha_{3}u_{4\mathbf{k}})(\alpha_{2}u_{1\mathbf{k}}+\alpha_{4}u_{3\mathbf{k}})^{\star}
\\
&=\sum_{in}\alpha_{1}\alpha_{2}(u_{1\mathbf{k}}u_{2\mathbf{k}}^{\star}-u_{2\mathbf{k}}u_{1\mathbf{k}}^{\star})+\alpha_{3}\alpha_{4}(u_{3\mathbf{k}}u_{4\mathbf{k}}^{\star}-u_{4\mathbf{k}}u_{3\mathbf{k}}^{\star})+\alpha_{1}\alpha_{4}(u_{1\mathbf{k}}u_{4\mathbf{k}}^{\star}-u_{2\mathbf{k}}u_{3\mathbf{k}}^{\star})+\alpha_{2}\alpha_{3}(u_{3\mathbf{k}}u_{2\mathbf{k}}^{\star}-u_{4\mathbf{k}}u_{1\mathbf{k}}^{\star})
\\
&=\sum_{in}\alpha_{1}\alpha_{4}(u_{1\mathbf{k}}u_{4\mathbf{k}}^{\star}-u_{2\mathbf{k}}u_{3\mathbf{k}}^{\star})+\alpha_{2}\alpha_{3}(u_{3\mathbf{k}}u_{2\mathbf{k}}^{\star}-u_{4\mathbf{k}}u_{1\mathbf{k}}^{\star}),
\end{aligned}
\end{equation}
where the first two terms in the second line are proportional to $\langle\sigma_{z}\rangle$ in different sectors, and should vanish due to time-reversal and inversion symmetries. Then, 
\begin{equation}
\label{eq:difference3}
\begin{aligned}
\Delta M_{f\mathbf{k}}&=\sum_{in}8\operatorname{Im}(A_{x}A_{y}^{\star})\alpha_{2}\alpha_{4}\operatorname{Re}(u_{2\mathbf{k}}^{\star}u_{4\mathbf{k}}-u_{1\mathbf{k}}^{\star}u_{3\mathbf{k}})
\\
&+4\operatorname{Re}\{A_{z}A_{y}^{\star}[\alpha_{1}\alpha_{4}(u_{1\mathbf{k}}u_{4\mathbf{k}}^{\star}-u_{2\mathbf{k}}u_{3\mathbf{k}}^{\star})+\alpha_{2}\alpha_{3}(u_{3\mathbf{k}}u_{2\mathbf{k}}^{\star}-u_{4\mathbf{k}}u_{1\mathbf{k}}^{\star})]\}.
\end{aligned}
\end{equation}
For Eq.~(1) in the main text, if the linear term dominates, the eigenstates are 
\begin{equation}
\label{eq:eigenstates}
\begin{aligned}
\ket{\psi_{1v}}=\frac{1}{\sqrt{2E(E-m)}}(E-m,0,0,k_{+})^{T},
\\
\ket{\psi_{2v}}=\frac{1}{\sqrt{2E(E-m)}}(0,E-m,k_{-},0)^{T},
\\
\ket{\psi_{1c}}=\frac{1}{\sqrt{2E(E+m)}}(E+m,0,0,k_{+})^{T},
\\
\ket{\psi_{2c}}=\frac{1}{\sqrt{2E(E+m)}}(0,E+m,k_{-},0)^{T},
\end{aligned}
\end{equation}
where $m=M+\lambda k_{z}^2$ and $E=\sqrt{k_{x}^2+k_{y}^2+m^2}$ and $k_{\pm}=k_{x}\pm i k_{y}$. Plugging Eq.~\eqref{eq:eigenstates} into Eq.~\eqref{eq:difference3}, one can get
\begin{equation}
\label{eq:difference4}
\begin{aligned}
\Delta M_{f\mathbf{k}}&=16A_{0}^2\cos\theta_{0}(\alpha_{1}\alpha_{4}+\alpha_{2}\alpha_{3})(E-m)k_{y}.
\end{aligned}
\end{equation}
for valence states, and 
\begin{equation}
\label{eq:difference5}
\begin{aligned}
\Delta M_{f\mathbf{k}}&=16A_{0}^2\cos\theta_{0}(\alpha_{1}\alpha_{4}+\alpha_{2}\alpha_{3})(E+m)k_{y}.
\end{aligned}
\end{equation}
for conduction states. One can directly see that $\Delta M_{f\mathbf{k}}$ has the same sign for conduction and valence states, but will change sign when $k_{y}$ changes sign. This matches the results of PtTe$_{2}$. 

On the other hand, if the cubic term dominates, then the eigenstates will be
\begin{equation}
\label{eq:eigenstates1}
\begin{aligned}
\ket{\psi_{1v}}=\frac{1}{\sqrt{2E(E-m)}}(0,E-m,0,i(k_{+}^{3}-k_{-}^{3}))^{T},
\\
\ket{\psi_{2v}}=\frac{1}{\sqrt{2E(E-m)}}(-(E-m),0,i(k_{+}^{3}-k_{-}^{3}),0)^{T},
\\
\ket{\psi_{1c}}=\frac{1}{\sqrt{2E(E+m)}}(0,-(E+m),0,i(k_{+}^{3}-k_{-}^{3}))^{T},
\\
\ket{\psi_{2c}}=\frac{1}{\sqrt{2E(E+m)}}(E+m,0,i(k_{+}^{3}-k_{-}^{3}),0)^{T},
\end{aligned}
\end{equation}
where $E=\sqrt{m^2+[i(k_{+}^{3}-k_{-}^{3})]^2}$. Then, 
\begin{equation}
\label{eq:difference6}
\begin{aligned}
\Delta M_{f\mathbf{k}}&=-16A_{0}^2\sin\theta_{0}\alpha_{2}\alpha_{4}(E-m)2k_{y}(k_{y}^2-3k_{x}^2)
\end{aligned}
\end{equation}
for valence states, and 
\begin{equation}
\label{eq:difference7}
\begin{aligned}
\Delta M_{f\mathbf{k}}&=16A_{0}^2\sin\theta_{0}\alpha_{2}\alpha_{4}(E+m)2k_{y}(k_{y}^2-3k_{x}^2).
\end{aligned}
\end{equation}
for conduction states. It is straightforward to see that $\Delta M_{f,\mathbf{k}}$ has different signs for conduction and valence states, and will change sign when $k_{y}$ changes sign. This matches the results of PdTe$_{2}$. 

\end{document}